\documentclass[journal]{IEEEtran}

\usepackage[utf8]{inputenc}
\usepackage{color}
\usepackage{array}
\usepackage{verbatim}
\usepackage{float}
\usepackage{amsmath}
\usepackage{amsthm}
\usepackage{amssymb}
\usepackage{graphicx}
\usepackage{longtable}
\usepackage{multirow}
\usepackage[unicode=true,
bookmarks=false,
breaklinks=false,pdfborder={0 0 1},colorlinks=false]
{hyperref}
\hypersetup{
	colorlinks,bookmarksopen,bookmarksnumbered,citecolor=blue,urlcolor=blue}
\usepackage{cite}

\floatstyle{ruled}
\newfloat{algorithm}{tbp}{loa}
\providecommand{\algorithmname}{Algorithm}
\floatname{algorithm}{\protect\algorithmname}

\makeatletter
\let\oldforeign@language\foreign@language
\DeclareRobustCommand{\foreign@language}[1]{%
	\lowercase{\oldforeign@language{#1}}}

\let\oldforeign@language\foreign@language
\DeclareRobustCommand{\foreign@language}[1]{%
	\lowercase{\oldforeign@language{#1}}}

\ifCLASSINFOpdf
\else
\fi

\newcommand{\MYfooter}{\smash{
		\hfil\parbox[t][\height][t]{\textwidth}{\centering
			\thepage}\hfil\hbox{}}}

\makeatletter

\def\ps@IEEEtitlepagestyle{%
	\def\@oddhead{\parbox[t][\height][t]{\textwidth}{\centering \scriptsize
			Personal use of this material is permitted. Permission from the author(s) and/or copyright holder(s), must be obtained for all other uses. Please contact us and provide details if you believe this document breaches copyrights.\\
			\noindent\makebox[\linewidth]{}
		}\hfil\hbox{}}%
	\def\@evenhead{\scriptsize\thepage \hfil \leftmark\mbox{}}%
	\def\@oddfoot{\parbox[t][\height][l]{\textwidth}{
			\vspace{-20pt}{\rule{\textwidth}{0.4pt}}\\ \footnotesize\underline{To cite this article:}
			{\bf{\textcolor{red}{H. A. Hashim, "Systematic Convergence of Nonlinear Stochastic Estimators on the Special Orthogonal Group SO(3)," International Journal of Robust and Nonlinear Control, vol. 30, no. 10, pp. 3848-3870, 2020.}}} doi: \href{https://doi.org/10.1002/RNC.4971}{10.1002/RNC.4971}\\
			\noindent\makebox[\linewidth]
		}\hfil\hbox{}}%
	\def\@evenfoot{\MYfooter}}

\makeatother
\newtheorem{defn}{Definition}

\newtheorem{lem}{Lemma}

\newtheorem{prop}{Proposition}
\newtheorem{thm}{Theorem}
\newtheorem{rem}{Remark}

\newtheorem{assum}{Assumption}

\begin{document}
	\bstctlcite{IEEEexample:BSTcontrol}

\title{Systematic Convergence of Nonlinear Stochastic Estimators on the Special Orthogonal Group SO(3)}

\author{Hashim A. Hashim
	\thanks{$^*$Corresponding author, H. A. Hashim is with the Department of Engineering and Applied Science, Thompson Rivers University, Kamloops, British Columbia, Canada, V2C-0C8, e-mail: hhashim@tru.ca.}
	
}


\markboth{}{Hashim \MakeLowercase{\textit{et al.}}: Nonlinear Pose Filters on SE(3) with Guaranteed Transient and Steady-state Performance}

\maketitle

\begin{abstract}
This paper introduces two novel nonlinear stochastic attitude estimators
developed on the Special Orthogonal Group $\mathbb{SO}\left(3\right)$
with the tracking error of the normalized Euclidean distance meeting
predefined transient and steady-state characteristics. The tracking
error is confined to initially start within a predetermined large
set such that the transient performance is guaranteed to obey dynamically
reducing boundaries and decrease smoothly and asymptotically to the
origin in probability from almost any initial condition. The proposed
estimators produce accurate attitude estimates with remarkable convergence
properties using measurements obtained from low-cost inertial measurement
units. The estimators proposed in continuous form are complemented
by their discrete versions for the implementation purposes. The quaternion representation of the proposed observers is provided. The simulation
results illustrate the effectiveness and robustness of the proposed
estimators against uncertain measurements and large initialization
error, whether in continuous or discrete form.
\end{abstract}

\begin{IEEEkeywords}
Attitude estimates, transient, steady-state error, nonlinear filter,
special orthogonal group, SO(3), stochastic system, stochastic differential equations,
Ito, Stratonovich, asymptotic stability, Wong-Zakai, inertial measurment unit, IMU, prescribed performance function.
\end{IEEEkeywords}

\IEEEpeerreviewmaketitle{}

\section{Introduction}

\IEEEPARstart{E}{Execution} of successful autonomous maneuvers requires accurate information
regarding the orientation of the vehicle. Nonetheless, the orientation,
commonly known as attitude, cannot be obtained directly. Instead,
attitude has to be acquired, using a set of measurements made on the
body-frame. These measurements are subject to unknown bias and noise
components, especially if they are supplied by low-cost inertial measurement
units (IMU). There are multiple ways to approach the problem of the
attitude estimation. For instance, the attitude can be established
algebraically using QUEST \cite{shuster1981three} and singular value
decomposition (SVD) \cite{markley1988attitude}. However the static
methods of estimation presented in \cite{shuster1981three} and \cite{markley1988attitude}
are characterized by poor results which differ significantly from
the true attitude \cite{hashim2018SO3Stochastic,hashim2018Conf1,mohamed2019filters,hashim2020AtiitudeSurvey}.

Historically, the attitude estimation problem has been tackled using
Gaussian filters, such as Kalman filter (KF) \cite{choukroun2006novel},
extended KF (EKF) \cite{lefferts1982kalman}, multiplicative EKF \cite{markley2003attitude}.
A complete survey of Gaussian attitude filters can be found in \cite{hashim2018SO3Stochastic}.
However, rapid development of low-cost IMU and natural non-linearity
of the attitude problem led to the proposal of multiple nonlinear
deterministic attitude estimators, for instance \cite{mahony2008nonlinear,zlotnik2017nonlinear,grip2012attitude,mahony2005complementary,hashim2019SO3Det}.
In fact, nonlinear deterministic attitude estimators received considerable
attention and have notable comparative advantages over Gaussian attitude
filters, namely, they are characterized by better tracking performance
and less computational power in comparison with Gaussian attitude
filters \cite{hashim2018SO3Stochastic,mohamed2019filters,mahony2008nonlinear,hashim2020AtiitudeSurvey}.
Nonlinear deterministic attitude estimators can be easily constructed
knowing a rate gyroscope measurement and two or more vector measurements.
A crucial part of the attitude estimator design is the selection of
the error function which has significant influence on the transient
performance and the steady-state error. The error function proposed
in \cite{mahony2005complementary} has been slightly modified in \cite{mahony2008nonlinear,grip2012attitude}.
However, the overall performance has not changed notably. The main
limitation of the error functions proposed in \cite{mahony2008nonlinear,grip2012attitude,mahony2005complementary}
is the slow convergence, in particular, when faced with large initial
attitude error. An alternative error function presented in \cite{zlotnik2017nonlinear,lee2012exponential}
allows for faster error convergence to the origin. Nevertheless, the
transient performance and steady-state error of the estimators in
\cite{zlotnik2017nonlinear,lee2012exponential} cannot be predicted,
and hence no systematic convergence is observed. In addition, nonlinear
deterministic estimators consider only constant bias attached to the
angular velocity measurements disregarding irregularity of the noise
behavior. Therefore, successful maneuvering applications, for instance
\cite{wang2018distributed,peng2019distributed,sofyali2019robust},
might not be achieved without robust attitude estimators.

The systematic convergence of the tracking error can be achieved by
enclosing the error to initially begin inside a predetermined large
set and diminish gradually and smoothly to a predetermined small set
\cite{bechlioulis2008robust}. This can be accomplished by transforming
the constrained error to its unconstrained shape, known by transformed
error. This transformation enables to maintain the error within dynamically
reducing sets enabling the attainment of the prescribed performance
measures. The systematic convergence with predefined measures has
been applied successfully in different applications, for instance
2 DOF planar robot \cite{bechlioulis2008robust}, unknown high-order
nonlinear networked systems \cite{hashim2017neuro}, event-triggered
control \cite{zheng2018event}.

Taking into consideration all of the above-mentioned challenges, two
robust nonlinear stochastic attitude estimators on the Special Orthogonal
Group $\mathbb{SO}\left(3\right)$ with predetermined transient as
well as steady-state measures have been proposed. As part of the proposed
approach, the attitude error is defined in terms of normalized Euclidean
distance. This error function is bound to start within a known large
set and decay in a systematic fashion to a given small set. As such,
the error function is constrained and as it converges to the origin,
the transformed error approaches zero in probability. The main contributions
of this work are as follow:
\begin{enumerate}
	\item[\textbf{1)}] The estimator design takes into consideration the unknown bias and
	random noise attached to angular velocity measurements.
	\item[\textbf{2)}] Taking into account the noise components allows to approach the attitude
	problem in stochastic sense, unlike \cite{mahony2008nonlinear,zlotnik2017nonlinear,grip2012attitude,mahony2005complementary,lee2012exponential}
	which disregarded the noise and tackled the problem in deterministic
	sense.
	\item[\textbf{3)}] The transformed error and normalized Euclidean distance of the attitude
	error are regulated to the origin in probability from almost any initial
	condition.
	\item[\textbf{4)}] The convergence of the estimators guarantees prescribed measures
	of transient and steady-state performance, unlike deterministic estimators
	in \cite{mahony2008nonlinear,zlotnik2017nonlinear,grip2012attitude,mahony2005complementary,lee2012exponential}.
	The proposed attitude estimators guarantee faster convergence properties
	owing to the dynamic behavior of the estimator gains. %
\end{enumerate}
The remainder of the paper is organized as follows: An overview of
$\mathbb{SO}\left(3\right)$ parameterization and mathematical notation
are presented in Section \ref{sec:SO3_PPF_STCH_Math-Notations}. In
Section \ref{sec:SO3_PPF_STCH_Problem-Formulation-in} the attitude
problem is framed in stochastic sense and the attitude error is formulated
in terms of prescribed performance. Nonlinear stochastic attitude
estimators with prescribed performance characteristics and the related
stability analysis are contained in Section \ref{sec:SO3PPF-Filters}.
The robustness of the proposed estimators is illustrated in Section
\ref{sec:SO3_PPF_STCH_Simulations}. Finally, Section \ref{sec:SO3_PPF_STCH_Conclusion}
summarizes the work.

\section{Notations and Preliminaries \label{sec:SO3_PPF_STCH_Math-Notations}}

In this paper, the set of non-negative real numbers is represented
by $\mathbb{R}_{+}$, $\mathbb{R}^{n}$ is the real $n$-dimensional
space, and $\mathbb{R}^{n\times m}$ is the real $n\times m$ dimensional
space. The set of integer numbers is represented by $\mathbb{N}$.
The Euclidean norm of $x\in\mathbb{R}^{n}$ is given by $||x||=\sqrt{x^{\top}x}$,
where superscript $^{\top}$ indicates transpose of a vector or a
matrix. $\lambda\left(\cdot\right)$ stands for a set of eigenvalues
of a matrix with $\underline{\lambda}\left(\cdot\right)$ being the
minimum eigenvalue within $\lambda\left(\cdot\right)$. $\mathcal{C}^{n}$
is a set of functions each of which is characterized by the $n$th
continuous partial derivative. $\mathbf{I}_{n}$ is an identity matrix
with $n$-by-$n$ dimensions. $\mathbb{E}\left[\cdot\right]$, $\mathbb{P}\left\{ \cdot\right\} $,
and $\exp\left(\cdot\right)$ signify expected value, probability,
and exponential of a component, respectively. Define the Special Orthogonal
Group as $\mathbb{SO}\left(3\right)$. Bearing in mind the notation
above, the attitude of a rigid-body is stated as a rotational matrix
$R$:
\[
\mathbb{SO}\left(3\right):=\left\{ \left.R\in\mathbb{R}^{3\times3}\right|R^{\top}R=RR^{\top}=\mathbf{I}_{3}\text{, }{\rm det}\left(R\right)=1\right\} 
\]
where ${\rm det\left(\cdot\right)}$ is the determinant of a matrix.
$\mathfrak{so}\left(3\right)$ is the Lie-algebra associated with
$\mathbb{SO}\left(3\right)$ defined by
\[
\mathfrak{so}\left(3\right):=\left\{ \left.\mathcal{X}\in\mathbb{R}^{3\times3}\right|\mathcal{X}^{\top}=-\mathcal{X}\right\} 
\]
with $\mathcal{X}$ being a skew-symmetric matrix. The map $\left[\cdot\right]_{\times}:\mathbb{R}^{3}\rightarrow\mathfrak{so}\left(3\right)$
is defined by
\[
\mathcal{X}=\left[x\right]_{\times}=\left[\begin{array}{ccc}
0 & -x_{3} & x_{2}\\
x_{3} & 0 & -x_{1}\\
-x_{2} & x_{1} & 0
\end{array}\right],\hspace{1em}x=\left[\begin{array}{c}
x_{1}\\
x_{2}\\
x_{3}
\end{array}\right]\in\mathbb{R}^{3}
\]
For any $x,y\in\mathbb{R}^{3}$, we have $\left[x\right]_{\times}y=x\times y$
with $\times$ being a cross product of the two vectors. The inverse
mapping of $\left[\cdot\right]_{\times}$ is defined by a vex operator,
which in turn can be expressed as $\mathbf{vex}:\mathfrak{so}\left(3\right)\rightarrow\mathbb{R}^{3}$
such that $\mathbf{vex}\left(\mathcal{X}\right)=x$, $\forall x\in\mathbb{R}^{3}$
and $\mathcal{X}\in\mathfrak{so}\left(3\right)$. Let $\boldsymbol{\mathcal{P}}_{a}$
be the anti-symmetric projection operator on the Lie-algebra of $\mathfrak{so}\left(3\right)$
\cite{murray1994mathematical}, given by $\boldsymbol{\mathcal{P}}_{a}:\mathbb{R}^{3\times3}\rightarrow\mathfrak{so}\left(3\right)$
such that
\[
\boldsymbol{\mathcal{P}}_{a}\left(\mathcal{Y}\right)=\frac{1}{2}\left(\mathcal{Y}-\mathcal{Y}^{\top}\right)\in\mathfrak{so}\left(3\right)
\]
where $\mathcal{Y}\in\mathbb{R}^{3\times3}$. Define the composition
mapping $\boldsymbol{\Upsilon}\left(\cdot\right)$ as
\begin{equation}
\boldsymbol{\Upsilon}\left(\mathcal{Y}\right)=\mathbf{vex}\left(\boldsymbol{\mathcal{P}}_{a}\left(\mathcal{Y}\right)\right)\in\mathbb{R}^{3},\hspace{1em}\forall\mathcal{Y}\in\mathbb{R}^{3\times3}\label{eq:SO3_PPF_STCH_VEX}
\end{equation}
where $\boldsymbol{\Upsilon}:=\mathbf{vex}\circ\boldsymbol{\mathcal{P}}_{a}$.
For a rotational matrix $R\in\mathbb{SO}\left(3\right)$, the normalized
Euclidean distance is defined by
\begin{equation}
||R||_{I}:=\frac{1}{4}{\rm Tr}\left\{ \mathbf{I}_{3}-R\right\} \label{eq:SO3_PPF_STCH_Ecul_Dist}
\end{equation}
with ${\rm Tr}\left\{ \cdot\right\} $ being trace of a matrix and
$||R||_{I}\in\left[0,1\right]$. The following identities will prove
useful in the subsequent derivations: 
\begin{align}
\left[\alpha\times\beta\right]_{\times}= & \beta\alpha^{\top}-\alpha\beta^{\top},\quad\alpha,\beta\in{\rm \mathbb{R}}^{3}\label{eq:SO3_PPF_STCH_Identity1}\\
\left[R\alpha\right]_{\times}= & R\left[\alpha\right]_{\times}R^{\top},\quad R\in\mathbb{SO}\left(3\right),\alpha\in\mathbb{R}^{3}\label{eq:SO3_PPF_STCH_Identity2}\\
\left[\alpha\right]_{\times}^{2}= & -||\alpha||^{2}\mathbf{I}_{3}+\alpha\alpha^{\top},\quad\alpha\in\mathbb{R}^{3}\label{eq:SO3_PPF_STCH_Identity3}\\
\left[A,B\right]= & AB-BA,\quad A,B\in\mathbb{R}^{3\times3}\label{eq:SO3_PPF_STCH_Identity8}\\
{\rm Tr}\left\{ \left[A,B\right]\right\} = & {\rm Tr}\left\{ AB-BA\right\} =0,\quad A,B\in\mathbb{R}^{3\times3}\label{eq:SO3_PPF_STCH_Identity5}\\
{\rm Tr}\left\{ B\left[\alpha\right]_{\times}\right\} = & 0,\quad B=B^{\top}\in\mathbb{R}^{3\times3},\alpha\in\mathbb{R}^{3}\label{eq:SO3_PPF_STCH_Identity6}
\end{align}
\begin{align}
{\rm Tr}\left\{ A\left[\alpha\right]_{\times}\right\} = & {\rm Tr}\left\{ \boldsymbol{\mathcal{P}}_{a}\left(A\right)\left[\alpha\right]_{\times}\right\} =-2\mathbf{vex}\left(\boldsymbol{\mathcal{P}}_{a}\left(A\right)\right)^{\top}\alpha,\nonumber \\
& \qquad A\in\mathbb{R}^{3\times3},\alpha\in\mathbb{R}^{3}\label{eq:SO3_PPF_STCH_Identity7}\\
B\left[\alpha\right]_{\times}+\left[\alpha\right]_{\times}B & ={\rm Tr}\left\{ B\right\} \left[\alpha\right]_{\times}-\left[B\alpha\right]_{\times},\nonumber \\
& \qquad B=B^{\top}\in\mathbb{R}^{3\times3},\alpha\in\mathbb{R}^{3}\label{eq:SO3_PPF_STCH_Identity4}
\end{align}
For a unit-axis $u\in\mathbb{R}^{3}$ rotating at a rotational angle
$\theta\in\mathbb{R}$ in a 2-sphere $\mathbb{S}^{2}$, the attitude
of a rigid-body can be established through the mapping of angle-axis
parameterization to $\mathbb{SO}\left(3\right)$ defined by $\mathcal{R}_{\theta}:\mathbb{R}\times\mathbb{R}^{3}\rightarrow\mathbb{SO}\left(3\right)$
such that \cite{shuster1993survey}
\begin{align}
\mathcal{R}_{\theta}\left(\theta,u\right) & ={\rm exp}\left(\theta\left[u\right]_{\times}\right)\nonumber \\
& =\mathbf{I}_{3}+\sin\left(\theta\right)\left[u\right]_{\times}+\left(1-\cos\left(\theta\right)\right)\left[u\right]_{\times}^{2}\label{eq:SO3_PPF_STCH_att_ang}
\end{align}
A more thorough overview of attitude mapping, important properties
and helpful notes can be found in \cite{hashim2019AtiitudeSurvey}.

\section{Problem Formulation with Prescribed Performance \label{sec:SO3_PPF_STCH_Problem-Formulation-in}}

This section aims to provide an overview of the body-frame and gyroscope
measurements in the attitude estimation context, and develop the attitude
problem in stochastic sense. Subsequently, the attitude problem is
reformulated to follow predefined measures of transient and steady-state
performance.

\subsection{Measurements and Attitude Kinematics in Stochastic Sense \label{subsec:SO3_PPF_STCH_Attitude-Kinematics}}

\begin{figure}[h!]
	\centering{}\includegraphics[scale=0.4]{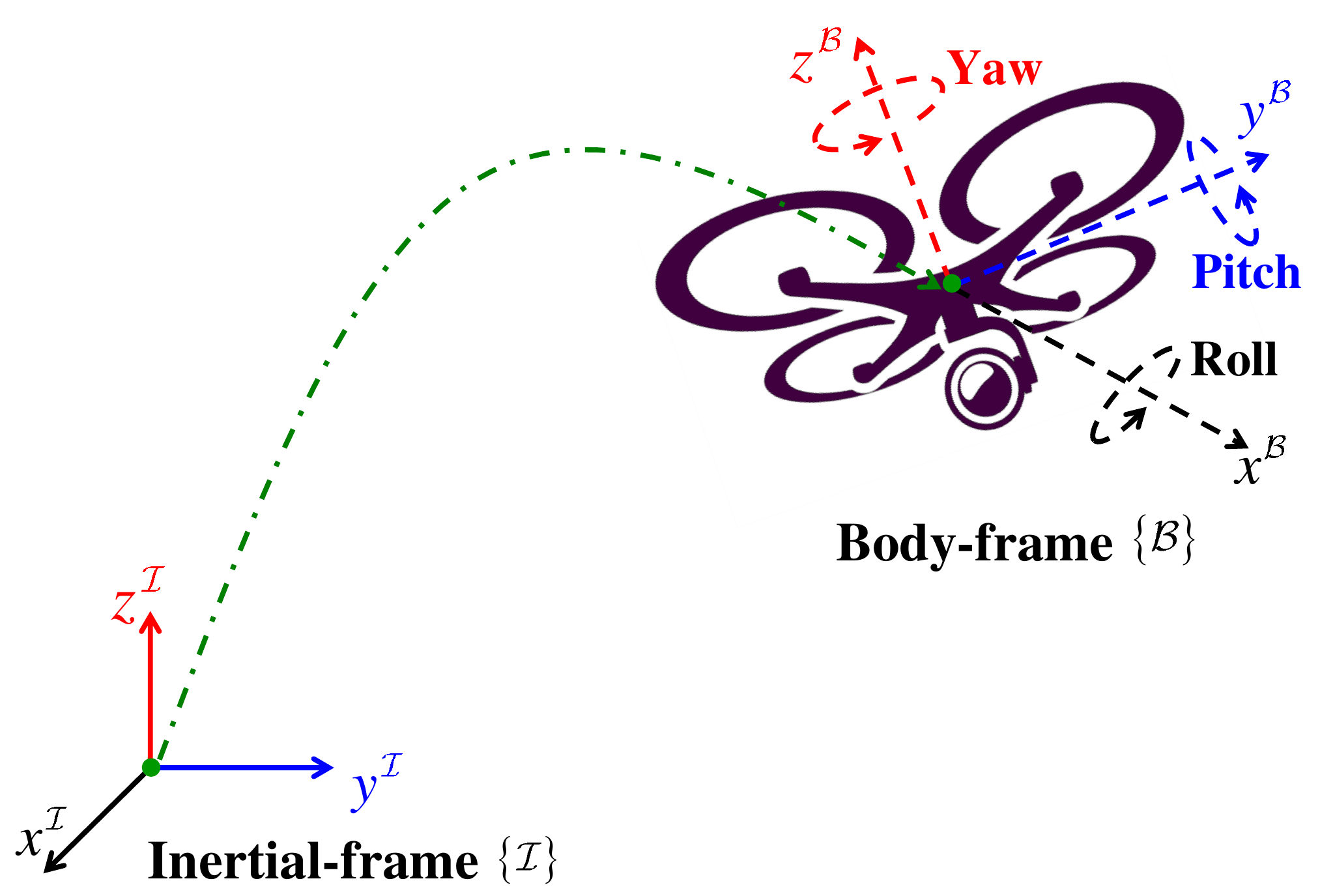}\caption{\label{fig:SO3_PPF_STCH_1}The orientation of a 3D rigid-body in body-frame
		relative to inertial-frame.}
\end{figure}

Let $R\in\mathbb{SO}\left(3\right)$ represent the relative orientation
of a rigid-body in the body-frame $\left\{ \mathcal{B}\right\} $
with respect to the inertial-frame $\left\{ \mathcal{I}\right\} $
as demonstrated in Fig. \ref{fig:SO3_PPF_STCH_1}. 

For the sake of clarity, the superscripts $\mathcal{I}$ and $\mathcal{B}$
refer to inertial-frame and body-frame vectors, respectively. Define
${\rm v}_{i}^{\mathcal{I}}$ as the $i$th known vector in the inertial-frame,
which is measured relative to the rigid-body fixed coordinate system
\begin{equation}
{\rm v}_{i}^{\mathcal{B}}=R^{\top}{\rm v}_{i}^{\mathcal{I}}+{\rm b}_{i}^{\mathcal{B}}+\omega_{i}^{\mathcal{B}}\in\mathbb{R}^{3}\label{eq:SO3_PPF_STCH_Vect_True}
\end{equation}
with ${\rm v}_{i}^{\mathcal{B}}$ being the $i$th body-frame measurement,
${\rm b}_{i}^{\mathcal{B}}$ being an unknown bias, and $\omega_{i}^{\mathcal{B}}$
denoting an unknown noise vector for all ${\rm v}_{i}^{\mathcal{I}},{\rm b}_{i}^{\mathcal{B}},\omega_{i}^{\mathcal{B}}\in\mathbb{R}^{3}$
and $i=1,2,\ldots,n$.
\begin{rem}
	\label{rem:Rem1}The attitude can be reconstructed if at least two
	instantaneously measured inertial-frame vectors as in \eqref{eq:SO3_PPF_STCH_Vect_True}
	are available ($n\geq2$). In case when $n=2$, the third body-frame
	and inertial-frame vectors can be obtained by ${\rm v}_{3}^{\mathcal{B}}={\rm v}_{1}^{\mathcal{B}}\times{\rm v}_{2}^{\mathcal{B}}$
	and ${\rm v}_{3}^{\mathcal{I}}={\rm v}_{1}^{\mathcal{I}}\times{\rm v}_{2}^{\mathcal{I}}$,
	respectively. 
\end{rem}
In accordance with Remark \ref{rem:Rem1}, in this work it is assumed
that $n\geq2$, and therefore, three non-collinear vectors can be
obtained using the expression in \eqref{eq:SO3_PPF_STCH_Vect_True}.
It is common practice to consider the normalized values of the inertial-frame
and body-frame vectors when calculating the attitude
\begin{equation}
\upsilon_{i}^{\mathcal{I}}=\frac{{\rm v}_{i}^{\mathcal{I}}}{||{\rm v}_{i}^{\mathcal{I}}||},\hspace{1em}\upsilon_{i}^{\mathcal{B}}=\frac{{\rm v}_{i}^{\mathcal{B}}}{||{\rm v}_{i}^{\mathcal{B}}||}\label{eq:SO3_PPF_STCH_Vector_norm}
\end{equation}
Thereby, the vectors defined in \eqref{eq:SO3_PPF_STCH_Vect_True}
will be normalized according to \eqref{eq:SO3_PPF_STCH_Vector_norm}
prior to estimating the attitude. For simplicity of stability analysis,
${\rm v}_{i}^{\mathcal{B}}$ is considered to be noise and bias free.
In the Simulation Section, however, ${\rm v}_{i}^{\mathcal{B}}$ is
regarded to be contaminated with noise and bias. The true attitude
kinematics are described by
\begin{equation}
\dot{R}=R\left[\Omega\right]_{\times}\label{eq:SO3_PPF_STCH_R_dynam}
\end{equation}
where $\Omega\in\left\{ \mathcal{B}\right\} $ denotes the true angular
velocity. Considering the normalized Euclidean distance of $R$ in
\eqref{eq:SO3_PPF_STCH_Ecul_Dist} and the identity in \eqref{eq:SO3_PPF_STCH_Identity7},
the kinematics in \eqref{eq:SO3_PPF_STCH_R_dynam} can be expressed
in terms of normalized Euclidean distance as
\begin{align}
\frac{d}{dt}||R||_{I} & =-\frac{1}{4}{\rm Tr}\left\{ \dot{R}\right\} \nonumber \\
& =-\frac{1}{4}{\rm Tr}\left\{ \boldsymbol{\mathcal{P}}_{a}\left(R\right)\left[\Omega\right]_{\times}\right\} \nonumber \\
& =\frac{1}{2}\boldsymbol{\Upsilon}\left(R\right)^{\top}\Omega\label{eq:SO3_PPF_STCH_NormR_dynam}
\end{align}
where $\boldsymbol{\Upsilon}\left(R\right)=\mathbf{vex}\left(\boldsymbol{\mathcal{P}}_{a}\left(R\right)\right)$.
The measurement of angular velocity is:
\begin{equation}
\Omega_{m}=\Omega+b+\omega\in\mathbb{R}^{3}\label{eq:SO3_PPF_STCH_Angular}
\end{equation}
where $b$ and $\omega$ represent the unknown constant bias and random
noise components attached to angular velocity measurements, respectively,
for all $b,\omega\in\mathbb{R}^{3}$. $\omega$ is a bounded Gaussian
random noise vector with zero mean. Derivative of any Gaussian process
is a Gaussian process, which allows $\omega$ to be expressed as a
function of Brownian motion process vector \cite{khasminskii1980stochastic,jazwinski2007stochastic}.
Assume that $\left\{ \omega,t\geq t_{0}\right\} $ is a vector process
of an independent Brownian motion process given by
\begin{equation}
\omega=\mathcal{Q}\frac{d\beta}{dt}\label{eq:SO3_PPF_STCH_noise}
\end{equation}
with $\beta\in\mathbb{R}^{3}$ and $\mathcal{Q}={\rm diag}\left(\mathcal{Q}_{1,1},\mathcal{Q}_{2,2},\mathcal{Q}_{3,3}\right)\in\mathbb{R}^{3\times3}$
being an unknown time-variant matrix where $\mathcal{Q}_{i,i}\in\mathbb{R}$
is a non-negative bounded component for all $i=1,2,3$. Also, ${\rm diag}\left(\cdot\right)$
denotes diagonal of a matrix. The covariance of the noise vector $\omega$
is defined by $\mathcal{Q}^{2}=\mathcal{Q}\mathcal{Q}^{\top}$. The
properties of the Brownian motion process are given as \cite{jazwinski2007stochastic,ito1984lectures,deng2001stabilization}
\[
\mathbb{P}\left\{ \beta\left(0\right)=0\right\} =1,\hspace{1em}\mathbb{E}\left[\beta\right]=0,\hspace{1em}\mathbb{E}\left[d\beta/dt\right]=0
\]
According to \eqref{eq:SO3_PPF_STCH_NormR_dynam}, \eqref{eq:SO3_PPF_STCH_Angular},
and \eqref{eq:SO3_PPF_STCH_noise}, the kinematics of the normalized
Euclidean distance in \eqref{eq:SO3_PPF_STCH_NormR_dynam} become
\begin{equation}
d||R||_{I}=\frac{1}{2}\boldsymbol{\Upsilon}\left(R\right)^{\top}\left(\left(\Omega_{m}-b\right)dt-\mathcal{Q}d\beta\right)\label{eq:SO3_PPF_STCH_NormR_Bias}
\end{equation}
Let us present Lemma \ref{Lemm:SO3_PPF_STCH_1} which will prove useful
in the subsequent estimator derivation. 
\begin{lem}
	\label{Lemm:SO3_PPF_STCH_1}Consider $R\in\mathbb{SO}\left(3\right)$,
	$M^{\mathcal{B}}=\left(M^{\mathcal{B}}\right)^{\top}\in\mathbb{R}^{3\times3}$,
	${\rm Tr}\left\{ M^{\mathcal{B}}\right\} =3$, and $\bar{\mathbf{M}}^{\mathcal{B}}={\rm Tr}\left\{ M^{\mathcal{B}}\right\} \mathbf{I}_{3}-M^{\mathcal{B}}$.
	Let the minimum singular value of $\bar{\mathbf{M}}^{\mathcal{B}}$
	be given by $\underline{\lambda}:=\underline{\lambda}\left(\bar{\mathbf{M}}^{\mathcal{B}}\right)$.
	Then, the following holds:
	\begin{equation}
	||\mathbf{vex}\left(\boldsymbol{\mathcal{P}}_{a}\left(R\right)\right)||^{2}=4\left(1-||R||_{I}\right)||R||_{I}\label{eq:SO3_PPF_STCH_lemm1_1}
	\end{equation}
	\begin{align}
	\frac{2}{\underline{\lambda}}\frac{||\mathbf{vex}\left(\boldsymbol{\mathcal{P}}_{a}\left(M^{\mathcal{B}}R\right)\right)||^{2}}{1+{\rm Tr}\left\{ \left(M^{\mathcal{B}}\right)^{-1}M^{\mathcal{B}}R\right\} } & \geq\left\Vert M^{\mathcal{B}}R\right\Vert _{I}\label{eq:SO3_PPF_STCH_lemm1_3}
	\end{align}
	\textbf{Proof. See \nameref{sec:SO3_PPF_STCH_AppendixA}.} 
\end{lem}
Let $\hat{R}$ be the estimate of the true attitude $R$. The objective
of any attitude estimator is to drive $\hat{R}\rightarrow R$ asymptotically
with fast convergence properties. Let the error in attitude from body-frame
to estimator-frame be given by
\begin{equation}
\tilde{R}=R^{\top}\hat{R}\label{eq:SO3_PPF_STCH_R_error}
\end{equation}
Also, consider the error in bias estimation to be defined by
\begin{align}
\tilde{b} & =b-\hat{b}\label{eq:SO3_PPF_STCH_b_tilde}
\end{align}

\subsection{Attitude Kinematics with Prescribed Performance \label{subsec:SO3_PPF_STCH_Prescribed-Performance}}

\begin{figure}[h!]
	\centering{}\includegraphics[scale=0.27]{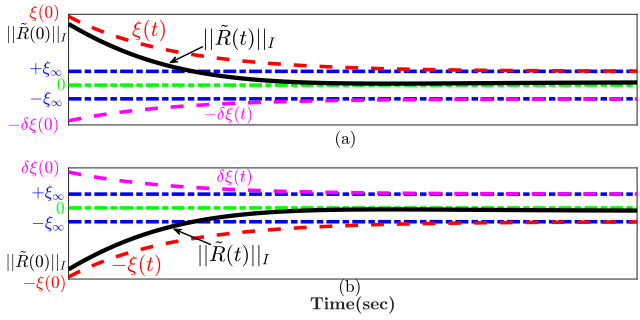} \caption{\label{fig:SO3_PPF_STCH_2}Systematic convergence of $||\tilde{R}\left(t\right)||_{I}$
		with PPF in accordance with (a) Eq. \eqref{eq:SO3_PPF_STCH_ePos};
		(b) Eq. \eqref{eq:SO3_PPF_STCH_eNeg}.}
\end{figure}

In this subsection the normalized Euclidean distance of the attitude
error $||\tilde{R}\left(t\right)||_{I}$ is reformulated to satisfy
predefined measures of transient and steady-state performance set
by the user. The objective of the reformulation is to force the error
$||\tilde{R}\left(t\right)||_{I}$ to initiate within a predefined
large set and reduce smoothly and systematically to a given small
set guided by a prescribed performance function (PPF) \cite{bechlioulis2008robust}.
Consider defining the PPF as $\xi\left(t\right)$ which is a positive
and time-decreasing function that satisfies $\xi:\mathbb{R}_{+}\to\mathbb{R}_{+}$
and $\lim\limits _{t\to\infty}\xi\left(t\right)=\xi_{\infty}>0$ such
that
\begin{equation}
\xi\left(t\right)=\left(\xi_{0}-\xi_{\infty}\right)\exp\left(-\ell t\right)+\xi_{\infty}\label{eq:SO3_PPF_STCH_Presc}
\end{equation}
with $\xi_{0}=\xi\left(0\right)$ being the upper bound of the predetermined
large set, and $\xi_{\infty}$ being the maximum value of the predetermined
small set, implying that the steady-state error is confined by $\pm\xi_{\infty}$.
Meanwhile, $\ell$ is a positive constant that regulates the convergence
rate of $\xi\left(t\right)$ from $\xi_{0}$ to $\xi_{\infty}$. In
order for the error $||\tilde{R}\left(t\right)||_{I}$ to follow the
dynamically decreasing boundaries of the PPF in \eqref{eq:SO3_PPF_STCH_Presc},
the following conditions should be met:
\begin{align}
-\delta\xi\left(t\right)<||\tilde{R}\left(t\right)||_{I}<\xi\left(t\right), & \text{ if }||\tilde{R}\left(0\right)||_{I}\geq0,\forall t\geq0\label{eq:SO3_PPF_STCH_ePos}\\
-\xi\left(t\right)<||\tilde{R}\left(t\right)||_{I}<\delta\xi\left(t\right), & \text{ if }||\tilde{R}\left(0\right)||_{I}<0,\forall t\geq0\label{eq:SO3_PPF_STCH_eNeg}
\end{align}
where $\delta$ satisfies $1\geq\delta\geq0$. The systematic convergence
of $||\tilde{R}\left(t\right)||_{I}$ guided by the dynamically decreasing
constraints of PPF in accordance with \eqref{eq:SO3_PPF_STCH_ePos}
and \eqref{eq:SO3_PPF_STCH_eNeg} is depicted in Fig. \ref{fig:SO3_PPF_STCH_2}.

\begin{rem}
	\label{rem3}Based on \eqref{eq:SO3_PPF_STCH_Ecul_Dist}, $||\tilde{R}\left(0\right)||_{I}\in[0,1]$
	, that being the case, the conditions in \eqref{eq:SO3_PPF_STCH_ePos}
	and Figure \ref{fig:SO3_PPF_STCH_2}.(a) are always fulfilled. Consequently,
	the steady-state error is bounded by $[0,+\xi_{\infty}]$, and $||\tilde{R}\left(t\right)||_{I}$
	is confined between $\xi\left(t\right)$ and $0$ as illustrated in
	Fig. \ref{fig:SO3_PPF_STCH_2}.(a).
\end{rem}
Let us define the error in normalized Euclidean distance as follows:
\begin{equation}
||\tilde{R}\left(t\right)||_{I}=\xi\left(t\right)\mathcal{Z}\left(\mathcal{E}\right)\label{eq:SO3_PPF_STCH_e_Trans}
\end{equation}
where $\xi\left(t\right)\in\mathbb{R}$ is specified in \eqref{eq:SO3_PPF_STCH_Presc},
$\mathcal{E}\in\mathbb{R}$ denotes the transformed error, and $\mathcal{Z}\left(\mathcal{E}\right)$
is a smooth function that fulfills:
\begin{assum}
	\label{Assum:SO3_PPF_STCH_1}Assume that function $\mathcal{Z}\left(\mathcal{E}\right)$
	satisfies the following three conditions \cite{bechlioulis2008robust}:
\begin{enumerate}
	\item[C 1)] $\mathcal{Z}\left(\mathcal{E}\right)$ is smooth and strictly increasing.
	\item[C 2)] $\mathcal{Z}\left(\mathcal{E}\right)$ behaves as follows: \\
	$-\underline{\delta}<\mathcal{Z}\left(\mathcal{E}\right)<\bar{\delta},{\rm \text{ if }}||\tilde{R}\left(0\right)||_{I}\geq0$\\
	with $\bar{\delta}$ and $\underline{\delta}$ being positive constants
	for all $\bar{\delta}\leq\underline{\delta}$. 
	\item[C 3)]  $\underset{\mathcal{E}\rightarrow+\infty}{\lim}\mathcal{Z}\left(\mathcal{E}\right)=\bar{\delta}$
	and $\underset{\mathcal{E}\rightarrow-\infty}{\lim}\mathcal{Z}\left(\mathcal{E}\right)=-\underline{\delta}$
	such that
	\begin{equation}
	\mathcal{Z}\left(\mathcal{E}\right)=\begin{cases}
	\frac{\bar{\delta}\exp\left(\mathcal{E}\right)-\underline{\delta}\exp\left(-\mathcal{E}\right)}{\exp\left(\mathcal{E}\right)+\exp\left(-\mathcal{E}\right)}, & \text{ if }||\tilde{R}\left(0\right)||_{I}\geq0\\
	\frac{\underline{\delta}\exp\left(\mathcal{E}\right)-\bar{\delta}\exp\left(-\mathcal{E}\right)}{\exp\left(\mathcal{E}\right)+\exp\left(-\mathcal{E}\right)}, & \text{ if }||\tilde{R}\left(0\right)||_{I}<0
	\end{cases}\label{eq:SO3_PPF_STCH_Smooth}
	\end{equation}
\end{enumerate}
\end{assum}
From \eqref{eq:SO3_PPF_STCH_Smooth} the transformed error can be
expressed by
\begin{equation}
\mathcal{E}\left(||\tilde{R}\left(t\right)||_{I},\xi\left(t\right)\right)=\mathcal{Z}^{-1}\left(\frac{||\tilde{R}\left(t\right)||_{I}}{\xi\left(t\right)}\right)\label{eq:SO3_PPF_STCH_Trans1}
\end{equation}
with $\mathcal{E}\in\mathbb{R}$, $\mathcal{Z}\in\mathbb{R}$, and
$\mathcal{Z}^{-1}\in\mathbb{R}$ being smooth functions. For simplicity,
define $\xi:=\xi\left(t\right)$, $||\tilde{R}||_{I}:=||\tilde{R}\left(t\right)||_{I}$,
and $\mathcal{E}:=\mathcal{E}\left(\cdot,\cdot\right)$. Combining
\eqref{eq:SO3_PPF_STCH_Smooth} and \eqref{eq:SO3_PPF_STCH_Trans1},
the transformed error can be expressed as follows:
\begin{equation}
\begin{aligned}\mathcal{E}= & \frac{1}{2}\text{ln}\frac{\underline{\delta}+||\tilde{R}||_{I}/\xi}{\bar{\delta}-||\tilde{R}||_{I}/\xi}\end{aligned}
\label{eq:SO3_PPF_STCH_trans3}
\end{equation}

\begin{rem}
	\label{rem:SO3_PPF_STCH_1} %
	{} The prescribed performance is achieved, when the transient and the
	steady-state performance of the tracking error is bounded by the dynamic
	boundaries of $\xi\left(t\right)$, and the transformed error $\mathcal{E}\left(t\right)$
	is bounded for all $t\geq0$.
\end{rem}
\begin{prop}
	\label{Prop:SO3_PPF_STCH_1}Let the normalized Euclidean distance
	error be $||\tilde{R}||_{I}$ defined according to \eqref{eq:SO3_PPF_STCH_Ecul_Dist}.
	Also, let the transformed error be expressed as in \eqref{eq:SO3_PPF_STCH_trans3}
	provided that $\underline{\delta}=\bar{\delta}$. Then, the following
	holds: 
\end{prop}
\begin{enumerate}
	\item[(i)] $\mathcal{E}>0$ for all $||\tilde{R}||_{I}\neq0$ and $\mathcal{E}=0$
	only at $||\tilde{R}||_{I}=0$. 
	\item[(ii)] The critical point of $\mathcal{E}$ coincides with $||\tilde{R}||_{I}=0$. 
	\item[(iii)] The only critical point of $\mathcal{E}$ is $\tilde{R}=\mathbf{I}_{3}$. 
\end{enumerate}
\textbf{Proof.} Provided that $\underline{\delta}=\bar{\delta}$ and
$||\tilde{R}||_{I}\leq\xi$, $\left(\underline{\delta}+||\tilde{R}||_{I}/\xi\right)/\left(\bar{\delta}-||\tilde{R}||_{I}/\xi\right)$
in \eqref{eq:SO3_PPF_STCH_trans3} is always greater than or equal
to 1. Therefore, it becomes evident that $\mathcal{E}>0\forall||\tilde{R}||_{I}\neq0$
and $\mathcal{E}=0$ at $||\tilde{R}||_{I}=0$ which justifies (i).
The definition of the normalized Euclidean distance in \eqref{eq:SO3_PPF_STCH_Ecul_Dist}
states that $||\tilde{R}||_{I}=0$ if and only if $\tilde{R}=\mathbf{I}_{3}$.
Hence, the only critical point of $\mathcal{E}$ coincides with $\tilde{R}=\mathbf{I}_{3}$
and $||\tilde{R}||_{I}=0$ which proves (ii) and (iii). Define the
following variable as
\begin{equation}
\begin{split}\mu(||\tilde{R}||_{I},\xi) & =\frac{1}{4\xi}\left(\frac{1}{\underline{\delta}+||\tilde{R}||_{I}/\xi}+\frac{1}{\bar{\delta}-||\tilde{R}||_{I}/\xi}\right)\end{split}
\label{eq:SO3_PPF_STCH_mu1}
\end{equation}
According to \eqref{eq:SO3_PPF_STCH_e_Trans} one has
\begin{align}
||\tilde{R}||_{I} & =\xi\mathcal{F}\left(\mathcal{E}\right)=\xi\frac{\bar{\delta}\exp\left(\mathcal{E}\right)-\underline{\delta}\exp\left(-\mathcal{E}\right)}{\exp\left(\mathcal{E}\right)+\exp\left(-\mathcal{E}\right)}\label{eq:SO3_PPF_STCH_e_Trans_final}
\end{align}
Therefore, from \eqref{eq:SO3_PPF_STCH_mu1} and \eqref{eq:SO3_PPF_STCH_e_Trans_final}
one may find that the expression in \eqref{eq:SO3_PPF_STCH_mu1} is
equivalent to
\begin{equation}
\mu(\mathcal{E},\xi)=\frac{\exp\left(2\mathcal{E}\right)+\exp\left(-2\mathcal{E}\right)+2}{8\xi\bar{\delta}}\label{eq:SO3_PPF_STCH_mu}
\end{equation}
It follows that the transformed error dynamics can be
\begin{align}
\dot{\mathcal{E}} & =2\mu\left(\frac{d}{dt}||\tilde{R}||_{I}-\frac{\dot{\xi}}{\xi}||\tilde{R}||_{I}\right)\label{eq:SO3_PPF_STCH_Trans_dot}
\end{align}
where $\mu:=\mu(\mathcal{E},\xi)$ which could also be expressed by
\begin{equation}
d\mathcal{E}=f(\mathcal{E},\tilde{b})dt+g\left(\mathcal{E}\right)\mathcal{Q}d\beta\label{eq:SO3_PPF_STCH_Trans_dot_Main}
\end{equation}
where both $f(\mathcal{E},\tilde{b})$ and $g\left(\mathcal{E}\right)$
are to be defined in Section \ref{sec:SO3PPF-Filters} with $g:\mathbb{R}\rightarrow\mathbb{R}^{1\times3}$
and $f:\mathbb{R}\times\mathbb{R}^{3}\rightarrow\mathbb{R}$. $g\left(\mathcal{E}\right)$
is locally Lipschitz in $\mathcal{E}$, and $f(\mathcal{E},\tilde{b})$
is locally Lipschitz in $\mathcal{E}$ and $\tilde{b}$ with $g\left(0\right)=\underline{\boldsymbol{0}}_{3}^{\top}$
and $f\left(0,\tilde{b}\right)=0$ for all $t\geq0$. Therefore, it
can be concluded that there exists a solution in the mean square sense
for the dynamic system in \eqref{eq:SO3_PPF_STCH_Trans_dot_Main}
for $t\in\left[t_{0},T\right]\forall t_{0}\leq T<\infty$ \cite{jazwinski2007stochastic,deng2001stabilization,deng1997stochastic}.
\begin{defn}
	\label{rem:Unstable-set}Define $\mathcal{S}\subseteq\mathbb{SO}\left(3\right)$
	as a non-attractive, forward invariant unstable set:
	\begin{equation}
	\mathcal{S}=\left\{ \left.\tilde{R}\left(0\right)\in\mathbb{SO}\left(3\right)\right|{\rm Tr}\{\tilde{R}\left(0\right)\}=-1\right\} \label{eq:SO3_PPF_STCH_SET}
	\end{equation}
	where the only three possible scenarios for $\tilde{R}\left(0\right)\in\mathcal{S}$
	are: $\tilde{R}\left(0\right)={\rm diag}(1,-1,-1)$, $\tilde{R}\left(0\right)={\rm diag}(-1,1,-1)$,
	and $\tilde{R}\left(0\right)={\rm diag}(-1,-1,1)$.
\end{defn}
For any $\mathcal{E}\left(t\right)\in\mathbb{R}$ that satisfies $t\neq t_{0}$
and $\tilde{R}\left(0\right)\notin\mathcal{S}$, $\mathcal{E}-\mathcal{E}_{0}$
is independent of $\left\{ \beta\left(\tau\right),\tau\geq t\right\} ,\forall t\in\left[t_{0},T\right]$
(Theorem 4.5 \cite{jazwinski2007stochastic}). With the objective
of achieving adaptive stabilization, let us define an unknown time-variant
covariance matrix $\mathcal{Q}^{2}$. For this purpose, let us also
specify the upper bound of $\mathcal{Q}^{2}$ as follows:
\begin{equation}
\sigma=\left[{\rm max}\left\{ \mathcal{Q}_{1,1}^{2}\right\} ,{\rm max}\left\{ \mathcal{Q}_{2,2}^{2}\right\} ,{\rm max}\left\{ \mathcal{Q}_{3,3}^{2}\right\} \right]^{\top}\in\mathbb{R}^{3}\label{eq:SO3_PPF_STCH_s_factor}
\end{equation}
where ${\rm max}\left\{ \cdot\right\} $ denotes the maximum value
of a component.
\begin{defn}
	\label{def:SO3_PPF_STCH_2} \cite{deng2001stabilization} For the
	stochastic dynamics in \eqref{eq:SO3_PPF_STCH_Trans_dot_Main}, and
	for a given potential function $V\left(\mathcal{E}\right)\in\mathcal{C}^{2}$,
	the differential operator $\mathcal{L}V$ is defined by
	\[
	\mathcal{L}V\left(\mathcal{E}\right)=V_{\mathcal{E}}^{\top}f(\mathcal{E},\tilde{b})+\frac{1}{2}{\rm Tr}\{g\left(\mathcal{E}\right)\mathcal{Q}^{2}g^{\top}\left(\mathcal{E}\right)V_{\mathcal{E}\mathcal{E}}\}
	\]
	where $V_{\mathcal{E}}=\partial V/\partial\mathcal{E}$ and $V_{\mathcal{E}\mathcal{E}}=\partial^{2}V/\partial\mathcal{E}^{2}$.
\end{defn}
\begin{lem}
	\label{lem:SO3_PPF_STCH_1} (Stochastic LaSalle Theorem \cite{krstic1998stabilization})
	Consider the stochastic dynamics in \eqref{eq:SO3_PPF_STCH_Trans_dot_Main}
	and suppose that there exists a positive definite function which is
	twice continuously differentiable $V\left(\mathcal{E}\right)\in\mathcal{C}^{2}$.
	Let $V\left(\mathcal{E}\right)$ be decrescent and radially unbounded,
	and suppose that there is a non-negative continuous class $\mathcal{K}$
	function $\boldsymbol{\mathcal{H}}\left(\mathcal{E}\right)\geq0$,
	such that $V\left(\mathcal{E}\right)$ of the stochastic kinematics
	in \eqref{eq:SO3_PPF_STCH_Trans_dot_Main} satisfies
	\begin{align}
	\mathcal{L}V\left(\mathcal{E}\right)= & V_{\mathcal{E}}^{\top}f\left(\mathcal{E}\right)+\frac{1}{2}{\rm Tr}\{g\left(\mathcal{E}\right)\mathcal{Q}^{2}g^{\top}\left(\mathcal{E}\right)V_{\mathcal{E}\mathcal{E}}\}\nonumber \\
	\leq & -\boldsymbol{\mathcal{H}}\left(\mathcal{E}\right),\hspace{1em}\forall\mathcal{E}\in\mathbb{R},t\geq0\label{eq:SO3_PPF_STCH_dVfunction_Lyap}
	\end{align}
	Thus, for $\mathcal{E}_{0}\in\mathbb{R}$ and the set in \eqref{eq:SO3_PPF_STCH_SET}
	$\tilde{R}\left(0\right)\notin\mathcal{S}$, there exists almost a
	unique strong solution on $\left[0,\infty\right)$ for the dynamic
	system in \eqref{eq:SO3_PPF_STCH_Trans_dot_Main}. Additionally, the
	equilibrium point $\mathcal{E}=0$ of the dynamic system in \eqref{eq:SO3_PPF_STCH_Trans_dot_Main}
	is almost globally asymptotically stable in probability such that
	\begin{equation}
	\mathbb{P}\{\lim_{t\rightarrow\infty}\boldsymbol{\mathcal{H}}\left(\mathcal{E}\left(t\right)\right)=0\}=1,\hspace{1em}\forall\tilde{R}\left(0\right)\notin\mathcal{S},\mathcal{E}\left(0\right)\in\mathbb{R}\label{eq:SO3_PPF_STCH_EVfunction_Lyap}
	\end{equation}
\end{lem}

\section{Nonlinear Stochastic Estimators on $\mathbb{SO}\left(3\right)$ with
	Prescribed Performance \label{sec:SO3PPF-Filters}}

This section introduces two nonlinear stochastic complimentary attitude
estimators on $\mathbb{SO}\left(3\right)$ which are capable of guiding
the normalized Euclidean distance error to follow the predetermined
transient and steady-state measures set by the user. $||\tilde{R}||_{I}$,
initially constrained, is transformed to an unconstrained error denoted
by $\mathcal{E}$. The first estimator is termed semi-direct nonlinear
stochastic estimator with prescribed performance as it demands attitude
reconstruction in addition to angular velocity measurements. The other
estimator is called direct nonlinear stochastic estimator with prescribed
performance since it uses a set of vector measurements in \eqref{eq:SO3_PPF_STCH_Vector_norm}
and \eqref{eq:SO3_PPF_STCH_Angular} avoiding the need for attitude
reconstruction. Let us define the error of the upper bound $\sigma$
as follows:
\begin{align}
\tilde{\sigma} & =\sigma-\hat{\sigma}\label{eq:SO3_PPF_STCH_s_tilde}
\end{align}
with $\hat{\sigma}$ being the estimate of $\sigma$.

\subsection{Semi-direct Stochastic Estimator with Prescribed Performance \label{subsec:SO3_PPF_STCH_Passive-Filter}}

Define $R_{y}$ as a reconstructed attitude of the rotational matrix
$R$. Consider the following estimator
\begin{align}
\mu= & \frac{\exp\left(2\mathcal{E}\right)+\exp\left(-2\mathcal{E}\right)+2}{8\xi\bar{\delta}}\label{eq:SO3_PPF_STCH_mu_Ry}\\
\dot{\hat{R}}= & \hat{R}\left[\Omega_{m}-\hat{b}-W\right]_{\times},\quad\hat{R}\left(0\right)=\hat{R}_{0}\label{eq:SO3_PPF_STCH_Rest_dot_Ry}\\
\dot{\hat{b}}= & \gamma_{1}\left(\mathcal{E}+1\right)\exp\left(\mathcal{E}\right)\mu\boldsymbol{\Upsilon}(\tilde{R})\label{eq:SO3_PPF_STCH_b_est_Ry}\\
\dot{\hat{\sigma}}= & \gamma_{2}\left(\mathcal{E}+2\right)\exp\left(\mathcal{E}\right)\mu^{2}{\rm diag}\left(\boldsymbol{\Upsilon}(\tilde{R})\right)\boldsymbol{\Upsilon}(\tilde{R})\label{eq:SO3_PPF_STCH_s_est_Ry}\\
W= & 2\frac{\mathcal{E}+2}{\mathcal{E}+1}\mu{\rm diag}\left(\boldsymbol{\Upsilon}(\tilde{R})\right)\hat{\sigma}+2\frac{k_{w}\mathcal{E}\mu-\dot{\xi}/4\xi}{1-||\tilde{R}||_{I}}\boldsymbol{\Upsilon}(\tilde{R})\label{eq:SO3_PPF_STCH_W_Ry}
\end{align}
with $\tilde{R}=R_{y}^{\top}\hat{R}$, $\boldsymbol{\Upsilon}(\tilde{R})={\rm vex}(\boldsymbol{\mathcal{P}}_{a}(\tilde{R}))$,
$\mathcal{E}$ being defined in \eqref{eq:SO3_PPF_STCH_trans3}, $\gamma_{1}$,
$\gamma_{2}$, and $k_{w}$ being positive constants, $||\tilde{R}||_{I}=\frac{1}{4}{\rm Tr}\{\mathbf{I}_{3}-\tilde{R}\}$
being defined in \eqref{eq:SO3_PPF_STCH_Ecul_Dist}, $\xi$ denoting
PPF defined in \eqref{eq:SO3_PPF_STCH_Presc}, and $\hat{b}$ and
$\hat{\sigma}$ being the estimates of $b$ and $\sigma$, respectively. The quaternion representation of the semi-direct proposed observer is given in \nameref{sec:SO3_PPF_STCH_AppendixB}.

\begin{thm}
	\textbf{\label{thm:SO3_PPF_STCH_1} }Consider the attitude dynamics
	in \eqref{eq:SO3_PPF_STCH_R_dynam}, vector measurements in \eqref{eq:SO3_PPF_STCH_Vect_True},
	and angular velocity measurements in \eqref{eq:SO3_PPF_STCH_Angular}
	coupled with the estimator in \eqref{eq:SO3_PPF_STCH_Rest_dot_Ry},
	\eqref{eq:SO3_PPF_STCH_b_est_Ry}, \eqref{eq:SO3_PPF_STCH_s_est_Ry},
	and \eqref{eq:SO3_PPF_STCH_W_Ry}. Assume that at least two body-frame
	non-collinear vectors are available for measurements. Recall the set
	defined in Definition \ref{rem:Unstable-set}. Thus, for attitude
	error $\tilde{R}\left(0\right)\notin\mathcal{S}$, and $\mathcal{E}\left(0\right)\in\mathbb{R}$,
	all the error signals are bounded, $\mathcal{E}$ asymptotically approaches
	the origin in probability, and $\tilde{R}$ asymptotically approaches
	$\mathbf{I}_{3}$ in probability. 
\end{thm}
Theorem \ref{thm:SO3_PPF_STCH_1} guarantees that the estimator kinematics
in \eqref{eq:SO3_PPF_STCH_Rest_dot_Ry}, \eqref{eq:SO3_PPF_STCH_b_est_Ry},
\eqref{eq:SO3_PPF_STCH_s_est_Ry}, and \eqref{eq:SO3_PPF_STCH_W_Ry}
are stable with $\mathcal{E}\left(t\right)$ being almost globally
asymptotically stable in probability. Since $\mathcal{E}\left(t\right)$
is bounded, $||\tilde{R}\left(t\right)||_{I}$ follows the dynamically
decrescent boundaries of the PPF introduced in \eqref{eq:SO3_PPF_STCH_Presc}.

\textbf{Proof. }Consider $\tilde{R}=R^{\top}\hat{R}$, $\tilde{b}=b-\hat{b}$,
and $\tilde{\sigma}=\sigma-\hat{\sigma}$ as in \eqref{eq:SO3_PPF_STCH_R_error},
\eqref{eq:SO3_PPF_STCH_b_tilde}, and \eqref{eq:SO3_PPF_STCH_s_tilde},
respectively. From \eqref{eq:SO3_PPF_STCH_R_dynam} and \eqref{eq:SO3_PPF_STCH_Rest_dot_Ry},
the error kinematics can be translated into
\begin{align}
d\tilde{R}= & R^{\top}\hat{R}\left[\Omega_{m}-\hat{b}-W\right]_{\times}dt+\left[\Omega\right]_{\times}^{\top}R^{\top}\hat{R}dt\nonumber \\
= & \left([\tilde{R},\left[\Omega\right]_{\times}]+\tilde{R}[\tilde{b}-W]_{\times}\right)dt+\tilde{R}\left[\mathcal{Q}d\beta\right]_{\times}\label{eq:SO3_PPF_STCH_Rtilde_dot}
\end{align}
where $[\tilde{R},\left[\Omega\right]_{\times}]=\tilde{R}\left[\Omega\right]_{\times}+\left[\Omega\right]_{\times}^{\top}\tilde{R}$
is the Lie bracket as defined in \eqref{eq:SO3_PPF_STCH_Identity8}.
In view of \eqref{eq:SO3_PPF_STCH_R_dynam} and \eqref{eq:SO3_PPF_STCH_NormR_dynam},
the error dynamics in \eqref{eq:SO3_PPF_STCH_Rtilde_dot} can be expressed
in the sense of normalized Euclidean distance as
\begin{align}
d||\tilde{R}||_{I}= & -\frac{1}{4}{\rm Tr}\left\{ \tilde{R}[\tilde{b}-W]_{\times}+\left[\tilde{R},\left[\Omega\right]_{\times}\right]\right\} dt\nonumber \\
& -\frac{1}{4}{\rm Tr}\left\{ \tilde{R}\left[\mathcal{Q}d\beta\right]_{\times}\right\} \nonumber \\
= & \frac{1}{2}\boldsymbol{\Upsilon}(\tilde{R})^{\top}\left((\tilde{b}-W)dt+\mathcal{Q}d\beta\right)\label{eq:SO3_PPF_STCH_NormRtilde_dot}
\end{align}
where ${\rm Tr}\{\tilde{R}[\tilde{b}]_{\times}\}=-2\boldsymbol{\Upsilon}(\tilde{R})^{\top}\tilde{b}$
as presented in identity \eqref{eq:SO3_PPF_STCH_Identity7}, and ${\rm Tr}\{[\tilde{R},\left[\Omega\right]_{\times}]\}=0$
as defined in identity \eqref{eq:SO3_PPF_STCH_Identity5}. From \eqref{eq:SO3_PPF_STCH_NormRtilde_dot},
the transformed error dynamics in incremental form are as follows:
\begin{align}
d\mathcal{E} & =f(\mathcal{E},\tilde{b})dt+g\left(\mathcal{E}\right)\mathcal{Q}d\beta\label{eq:SO3_PPF_STCH_Trans_Ry_ITO}
\end{align}
where $f(\mathcal{E},\tilde{b})=2\mu\left(\frac{1}{2}\boldsymbol{\Upsilon}(\tilde{R})^{\top}(\tilde{b}-W)-\frac{\dot{\xi}}{\xi}||\tilde{R}||_{I}\right)$
and $g\left(\mathcal{E}\right)=\mu\boldsymbol{\Upsilon}(\tilde{R})^{\top}$.
Ito \cite{khasminskii1980stochastic,ito1984lectures} and Stratonovich
\cite{stratonovich1967topics} are the two most commonly used methods
of stochastic integral representation. These two approaches have the
following two advantages: $\beta\left(t\right)$ and $d\beta/dt$
are continuous and Lipschitz, and their mean square exists. However,
the Ito integral does not obey the chain rule and can be a good fit
for measurements corrupted only by white noise signals. Stratonovich,
on the other hand, is a well-defined Riemann integral, and as opposed
to Ito, has a continuous partial derivative with respect to $\beta$
which means that it follows the chain rule. Additionally, Stratonovich
is suitable for both white and colored noise signals \cite{hashim2018SO3Stochastic,jazwinski2007stochastic,stratonovich1967topics}.
Suppose that the dynamics in \eqref{eq:SO3_PPF_STCH_Trans_Ry_ITO}
are defined in the sense of Stratonovich \cite{stratonovich1967topics}.
Thereby, the transformation of \eqref{eq:SO3_PPF_STCH_Trans_Ry_ITO}
from Stratonovich to Ito is given by
\begin{align}
d\mathcal{E} & =\mathcal{F}(\mathcal{E},\tilde{b})dt+g\left(\mathcal{E}\right)\mathcal{Q}d\beta\label{eq:SO3_PPF_STCH_Trans_Ry_STRAT}
\end{align}
such that
\[
\mathcal{F}(\mathcal{E},\tilde{b})=f(\mathcal{E},\tilde{b})+\boldsymbol{\mathcal{W}}\left(\mathcal{E}\right)
\]
where $\boldsymbol{\mathcal{W}}\left(\mathcal{E}\right)$ is the Wong-Zakai
factor introduced to allow for the transition from Stratonovich to
Ito and is given by \cite{jazwinski2007stochastic,stratonovich1967topics}
\begin{align}
\boldsymbol{\mathcal{W}}\left(\mathcal{E}\right) & =g\left(\mathcal{E}\right)\frac{\mathcal{Q}\mathcal{Q}^{\top}}{2}\frac{\partial g\left(\mathcal{E}\right)}{\partial\mathcal{E}}\nonumber \\
& =\frac{\boldsymbol{\Upsilon}(\tilde{R})^{\top}\mathcal{Q}^{2}\boldsymbol{\Upsilon}(\tilde{R})}{2}\frac{\partial\mu}{\partial\mathcal{E}}\mu\nonumber \\
& =\frac{\exp\left(2\mathcal{E}\right)-\exp\left(-2\mathcal{E}\right)}{8\xi\bar{\delta}}\mu\boldsymbol{\Upsilon}(\tilde{R})^{\top}\mathcal{Q}^{2}\boldsymbol{\Upsilon}(\tilde{R})\label{eq:SO3_PPF_STCH_Wong_STRAT}
\end{align}

\begin{rem}
	\label{rem:Wong-Zakai-factor}\cite{jazwinski2007stochastic,stratonovich1967topics}
	The Wong-Zakai factor introduced in \eqref{eq:SO3_PPF_STCH_Trans_Ry_STRAT}
	has a prominent role in attaining $\mathbb{E}\left[\mathcal{E}\right]=\mathbb{E}\left[\int_{t_{0}}^{t}f(\mathcal{E}\left(\tau\right),\tilde{b}\left(\tau\right))d\tau\right]$
	whether $d\beta/dt$ has a zero mean or not.
\end{rem}
Consider the following candidate Lyapunov function
\begin{align}
V(\mathcal{E},\tilde{b},\tilde{\sigma}) & =\mathcal{E}\exp\left(\mathcal{E}\right)+\frac{1}{2\gamma_{1}}||\tilde{b}||^{2}+\frac{1}{\gamma_{2}}||\tilde{\sigma}||^{2}\label{eq:SO3_PPF_STCH_V_Ry}
\end{align}
For $V:=V(\mathcal{E},\tilde{b},\tilde{\sigma})$ the first and second
partial derivatives of \eqref{eq:SO3_PPF_STCH_V_Ry} with respect
to $\mathcal{E}$ are as follows
\begin{align}
V_{\mathcal{E}}=\frac{\partial V}{\partial\mathcal{E}} & =\left(\mathcal{E}+1\right)\exp\left(\mathcal{E}\right)\label{eq:SO3_PPF_STCH_Vv_Ry}\\
V_{\mathcal{E}\mathcal{E}}=\frac{\partial^{2}V}{\partial\mathcal{E}^{2}} & =\left(\mathcal{E}+2\right)\exp\left(\mathcal{E}\right)\label{eq:SO3_PPF_STCH_Vvv_Ry}
\end{align}
Consider the differential operator $\mathcal{L}V$ in Definition \ref{def:SO3_PPF_STCH_2}.
From \eqref{eq:SO3_PPF_STCH_V_Ry}, \eqref{eq:SO3_PPF_STCH_Trans_Ry_STRAT},
\eqref{eq:SO3_PPF_STCH_Vv_Ry}, and \eqref{eq:SO3_PPF_STCH_Vvv_Ry},
$\mathcal{L}V$ is equivalent to
\begin{align}
\mathcal{L}V= & V_{\mathcal{E}}^{\top}\mathcal{F}(\mathcal{E},\tilde{b})+\frac{1}{2}{\rm Tr}\{g\left(\mathcal{E}\right)\mathcal{Q}^{2}g^{\top}\left(\mathcal{E}\right)V_{\mathcal{E}\mathcal{E}}\}\nonumber \\
& -\frac{1}{\gamma_{1}}\tilde{b}^{\top}\dot{\hat{b}}-\frac{2}{\gamma_{2}}\tilde{\sigma}^{\top}\dot{\hat{\sigma}}\nonumber \\
= & \left(\mathcal{E}+1\right)\exp\left(\mathcal{E}\right)\mu\left(\boldsymbol{\Upsilon}(\tilde{R})^{\top}(\tilde{b}-W)-2\frac{\dot{\xi}}{\xi}||\tilde{R}||_{I}\right.\nonumber \\
& \hspace{2em}\left.+\frac{\exp\left(2\mathcal{E}\right)-\exp\left(-2\mathcal{E}\right)}{8\xi\bar{\delta}}\boldsymbol{\Upsilon}(\tilde{R})^{\top}\mathcal{Q}^{2}\boldsymbol{\Upsilon}(\tilde{R})\right)\nonumber \\
& +\frac{1}{2}\left(\mathcal{E}+2\right)\exp\left(\mathcal{E}\right)\mu^{2}\boldsymbol{\Upsilon}(\tilde{R})^{\top}\mathcal{Q}^{2}\boldsymbol{\Upsilon}(\tilde{R})\nonumber \\
& -\frac{1}{\gamma_{1}}\tilde{b}^{\top}\dot{\hat{b}}-\frac{2}{\gamma_{2}}\tilde{\sigma}^{\top}\dot{\hat{\sigma}}\label{eq:SO3_PPF_STCH_Vdot_Ry_1}
\end{align}
From property (ii) and (iii) of Proposition \ref{Prop:SO3_PPF_STCH_1},
$\mathcal{E}>0\forall||\tilde{R}||_{I}\neq0$ and $\mathcal{E}=0$
only at $||\tilde{R}||_{I}=0$. From \eqref{eq:SO3_PPF_STCH_mu} and
\eqref{eq:SO3_PPF_STCH_mu_Ry}, one has $\mu>0\forall t\geq0$. Thus,
one has $\mathcal{E}\exp\left(\mathcal{E}\right)\geq0$, $\mu>0$
and $\boldsymbol{\Upsilon}(\tilde{R})^{\top}\mathcal{Q}^{2}\boldsymbol{\Upsilon}(\tilde{R})\geq0$
for all $t\geq0$. Also, it becomes apparent that $\mu>\frac{\exp\left(2\mathcal{E}\right)-\exp\left(-2\mathcal{E}\right)}{8\xi\bar{\delta}}$.
As such, the differential operator in \eqref{eq:SO3_PPF_STCH_Vdot_Ry_1}
can be written in inequality form as
\begin{align}
\mathcal{L}V\leq & 2\left(\mathcal{E}+1\right)\exp\left(\mathcal{E}\right)\mu\left(\frac{1}{2}\boldsymbol{\Upsilon}(\tilde{R})^{\top}(\tilde{b}-W)-\frac{\dot{\xi}}{\xi}||\tilde{R}||_{I}\right)\nonumber \\
& +2\left(\mathcal{E}+2\right)\exp\left(\mathcal{E}\right)\mu^{2}\boldsymbol{\Upsilon}(\tilde{R})^{\top}{\rm diag}\left(\boldsymbol{\Upsilon}(\tilde{R})\right)\sigma\nonumber \\
& -\frac{1}{\gamma_{1}}\tilde{b}^{\top}\dot{\hat{b}}-\frac{2}{\gamma_{2}}\tilde{\sigma}^{\top}\dot{\hat{\sigma}}\label{eq:SO3_PPF_STCH_Vdot_Ry_2}
\end{align}
where $\sigma$ is defined in \eqref{eq:SO3_PPF_STCH_s_factor}. Directly
substituting $\dot{\hat{b}}$, $\dot{\hat{\sigma}}$, and $W$ with
their definitions in \eqref{eq:SO3_PPF_STCH_b_est_Ry}, \eqref{eq:SO3_PPF_STCH_s_est_Ry},
and \eqref{eq:SO3_PPF_STCH_W_Ry}, respectively, and considering $||\boldsymbol{\Upsilon}(\tilde{R})||^{2}=4(1-||\tilde{R}||_{I})||\tilde{R}||_{I}$
as defined in \eqref{eq:SO3_PPF_STCH_lemm1_1}, the inequality in
\eqref{eq:SO3_PPF_STCH_Vdot_Ry_2} becomes
\begin{align}
\mathcal{L}V\leq & -4k_{w}||\tilde{R}||_{I}\mu^{2}\left(\mathcal{E}+1\right)\mathcal{E}\exp\left(\mathcal{E}\right)\label{eq:SO3_PPF_STCH_Vdot_Ry_3}
\end{align}
Since $\mu>0\forall t\geq0$, from \eqref{eq:SO3_PPF_STCH_e_Trans_final}
the inequality in \eqref{eq:SO3_PPF_STCH_Vdot_Ry_3} is equivalent
to
\begin{align}
\mathcal{L}V\leq & -4\bar{\delta}k_{w}\xi\mu^{2}\left(\mathcal{E}+1\right)\mathcal{E}\frac{\exp\left(\mathcal{E}\right)-\exp\left(-\mathcal{E}\right)}{\exp\left(\mathcal{E}\right)+\exp\left(-\mathcal{E}\right)}\exp\left(\mathcal{E}\right)\label{eq:SO3_PPF_STCH_Vdot_Ry_Final}
\end{align}
with $\bar{\delta}>0$, $k_{w}>0$, and $\xi:\mathbb{R}_{+}\to\mathbb{R}_{+}$.
As stated by property (ii) and (iii) of Proposition \ref{Prop:SO3_PPF_STCH_1},
$\mathcal{E}>0\forall||\tilde{R}||_{I}\neq0$ and $\mathcal{E}=0$
only at $||\tilde{R}||_{I}=0$ for $\tilde{R}\left(0\right)\notin\mathcal{S}$
and $\mathcal{E}\left(0\right)\in\mathbb{R}$. According to the fact
that $\mathcal{L}V$ is bounded and $V$ is radially unbounded for
any $\tilde{R}\left(0\right)\notin\mathcal{S}$ and $\mathcal{E}\left(0\right)\in\mathbb{R}$,
there exists a unique strong solution for the stochastic system in
\eqref{eq:SO3_PPF_STCH_Trans_Ry_STRAT} with probability of one \cite{khasminskii1980stochastic}.
From property (ii) and (iii) of Proposition \ref{Prop:SO3_PPF_STCH_1},
$\mathcal{E}=0$ implies that $\tilde{R}=\mathbf{I}_{3}$. Thus, based
on the result presented in \eqref{eq:SO3_PPF_STCH_Vdot_Ry_Final},
$\mathcal{E}=0$ and $\tilde{R}=\mathbf{I}_{3}$ are independent of
the unknown noise and bias attached to the angular velocity measurements.
Therefore, on the basis of the stochastic LaSalle Theorem \cite{khasminskii1980stochastic},
it can be concluded that $\mathcal{E}=0$ is almost globally stable
in probability with $\mathcal{E}\left(t\right)$ being regulated asymptotically
to the origin in probability for all $\tilde{R}\left(0\right)\notin\mathcal{S}$
and $\mathcal{E}\left(0\right)\in\mathbb{R}$ which implies that $\mathbb{P}\{\lim_{t\rightarrow\infty}\tilde{R}=\mathbf{I}_{3}\}=1$
for all $\tilde{R}\left(0\right)\notin\mathcal{S}$ and $\mathcal{E}\left(0\right)\in\mathbb{R}$.
Moreover, it can be noticed that the estimation of $b$ and $\sigma$
is achievable and has a finite limit in probability (\cite{deng2001stabilization},
Theorem 3.1).

\subsection{Direct Stochastic Estimator with Prescribed Performance\label{subsec:SO3_PPF_STCH_Explicit-Filter}}

Let us define $R_{y}$ as the reconstructed attitude of the true $R$.
The stochastic estimator introduced in the above subsection requires
attitude reconstruction. Several methods have been proposed to obtain
static estimation of the attitude, such as QUEST \cite{shuster1981three}
and SVD \cite{markley1988attitude}. Nonetheless, the aforementioned
methods add complexity to the estimator design process \cite{mahony2008nonlinear,hamel2006attitude}.
In an effort to simplify the estimator design, this section presents
a novel nonlinear stochastic attitude estimator with prescribed performance.
The merit of the proposed estimator consists in its ability to receive
direct input from the measurement units bypassing attitude reconstruction.
Consider the normalized inertial and body-frame vectors in \eqref{eq:SO3_PPF_STCH_Vector_norm}
for $i=1,\ldots,n$ and define
\begin{align}
M^{\mathcal{I}} & =\left(M^{\mathcal{I}}\right)^{\top}=\sum_{i=1}^{n}s_{i}\upsilon_{i}^{\mathcal{I}}\left(\upsilon_{i}^{\mathcal{I}}\right)^{\top}\nonumber \\
M^{\mathcal{B}} & =\left(M^{\mathcal{B}}\right)^{\top}=\sum_{i=1}^{n}s_{i}\upsilon_{i}^{\mathcal{B}}\left(\upsilon_{i}^{\mathcal{B}}\right)^{\top}=R^{\top}M^{\mathcal{I}}R\label{eq:SO3_PPF_STCH_MB_MI}
\end{align}
with $s_{i}>0$ representing the confidence level of the $i$th sensor
measurement. In this work, $s_{i}$ has been selected to satisfy $\sum_{i=1}^{n}s_{i}=3$.
It becomes apparent that $M^{\mathcal{I}}$ and $M^{\mathcal{B}}$
are symmetric matrices, and therefore suppose that at least two non-collinear
inertial-frame and body-frame vectors are available at every time
instant. In case when $n=2$, the third vector is defined using the
cross product of the two known vectors as stated in Subsection \ref{subsec:SO3_PPF_STCH_Attitude-Kinematics}.
As such, $M^{\mathcal{B}}$ is nonsingular with ${\rm rank}(M^{\mathcal{B}})=3$
at every time instant. In addition, the three eigenvalues of the symmetric
matrix $M^{\mathcal{B}}$ are greater than zero \cite{bullo2004geometric}.
Let us introduce $\bar{\mathbf{M}}^{\mathcal{B}}={\rm Tr}\{M^{\mathcal{B}}\}\mathbf{I}_{3}-M^{\mathcal{B}}\in\mathbb{R}^{3\times3}$,
given that $n\geq2$. Accordingly, the following statements are met
(\cite{bullo2004geometric} page. 553):  
\begin{enumerate}
	\item The matrix $\bar{\mathbf{M}}^{\mathcal{B}}$ is symmetric and positive-definite.
	
	\item The eigenvectors of $M^{\mathcal{B}}$ have values similar to those
	of the eigenvectors of $\bar{\mathbf{M}}^{\mathcal{B}}$.  
	\item Consider the three eigenvalues of $M^{\mathcal{B}}$ stated as $\lambda(M^{\mathcal{B}})=\{\lambda_{1},\lambda_{2},\lambda_{3}\}$.
	It follows that $\lambda(\bar{\mathbf{M}}^{\mathcal{B}})=\{\lambda_{3}+\lambda_{2},\lambda_{3}+\lambda_{1},\lambda_{2}+\lambda_{1}\}$
	with $\underline{\lambda}(\bar{\mathbf{M}}^{\mathcal{B}})>0$ being
	the minimum eigenvalue of $\bar{\mathbf{M}}^{\mathcal{B}}$.  
\end{enumerate}
In the remainder of this subsection it is supposed that $n\geq2$
such that ${\rm rank}(M^{\mathcal{B}})=3$. Define the estimate of
$\upsilon_{i}^{\mathcal{B}}$ by
\begin{equation}
\hat{\upsilon}_{i}^{\mathcal{B}}=\hat{R}^{\top}\upsilon_{i}^{\mathcal{I}}\label{eq:SO3_PPF_STCH_vB_hat}
\end{equation}
In order to avoid attitude reconstruction, it is necessary to introduce
a set of expressions in terms of vector measurements. With the aid
of identity \eqref{eq:SO3_PPF_STCH_Identity1}, one obtains
\begin{align*}
\left[\sum_{i=1}^{n}\frac{s_{i}}{2}\hat{\upsilon}_{i}^{\mathcal{B}}\times\upsilon_{i}^{\mathcal{B}}\right]_{\times} & =\sum_{i=1}^{n}\frac{s_{i}}{2}\left(\upsilon_{i}^{\mathcal{B}}\left(\hat{\upsilon}_{i}^{\mathcal{B}}\right)^{\top}-\hat{\upsilon}_{i}^{\mathcal{B}}\left(\upsilon_{i}^{\mathcal{B}}\right)^{\top}\right)\\
& =\frac{1}{2}R^{\top}M^{\mathcal{I}}R\tilde{R}-\frac{1}{2}\tilde{R}^{\top}R^{\top}M^{\mathcal{I}}R\\
& =\boldsymbol{\mathcal{P}}_{a}(M^{\mathcal{B}}\tilde{R})
\end{align*}
such that
\begin{equation}
\mathbf{vex}\left(\boldsymbol{\mathcal{P}}_{a}(M^{\mathcal{B}}\tilde{R})\right)=\sum_{i=1}^{n}\frac{s_{i}}{2}\hat{\upsilon}_{i}^{\mathcal{B}}\times\upsilon_{i}^{\mathcal{B}}\label{eq:SO3_PPF_STCH_VEX_VM}
\end{equation}
It is straight forward to find
\begin{align}
||M^{\mathcal{B}}\tilde{R}||_{I} & =\frac{1}{4}{\rm Tr}\{\mathbf{I}_{3}-M^{\mathcal{B}}\tilde{R}\}\nonumber \\
& =\frac{1}{4}{\rm Tr}\left\{ \mathbf{I}_{3}-\sum_{i=1}^{n}s_{i}\upsilon_{i}^{\mathcal{B}}\left(\hat{\upsilon}_{i}^{\mathcal{B}}\right)^{\top}\right\} \nonumber \\
& =\frac{1}{4}\sum_{i=1}^{n}s_{i}\left(1-\left(\hat{\upsilon}_{i}^{\mathcal{B}}\right)^{\top}\upsilon_{i}^{\mathcal{B}}\right)\label{eq:SO3_PPF_STCH_RI_VM}
\end{align}
Consider the following variable
\begin{align}
\boldsymbol{\mathcal{J}}(M^{\mathcal{B}},\tilde{R}) & ={\rm Tr}\left\{ \left(M^{\mathcal{B}}\right)^{-1}M^{\mathcal{B}}\tilde{R}\right\} \nonumber \\
& ={\rm Tr}\left\{ \left(\sum_{i=1}^{n}s_{i}\upsilon_{i}^{\mathcal{B}}\left(\upsilon_{i}^{\mathcal{B}}\right)^{\top}\right)^{-1}\sum_{i=1}^{n}s_{i}\upsilon_{i}^{\mathcal{B}}\left(\hat{\upsilon}_{i}^{\mathcal{B}}\right)^{\top}\right\} \label{eq:SO3_PPF_STCH_Gamma}
\end{align}
In the subsequent estimator derivations the variables $\mathbf{vex}(\boldsymbol{\mathcal{P}}_{a}(M^{\mathcal{B}}\tilde{R}))$,
$||M^{\mathcal{B}}\tilde{R}||_{I}$, and $\boldsymbol{\mathcal{J}}(M^{\mathcal{B}},\tilde{R})$
are going to be obtained through a set of vector measurements as defined
in \eqref{eq:SO3_PPF_STCH_VEX_VM}, \eqref{eq:SO3_PPF_STCH_RI_VM},
and \eqref{eq:SO3_PPF_STCH_Gamma}, respectively. Recall the discussion
presented in Subsection \ref{subsec:SO3_PPF_STCH_Prescribed-Performance}:
every $||\tilde{R}||_{I}$ is replaced by $||M^{\mathcal{B}}\tilde{R}||_{I}$
such that $\mathcal{E}:=\mathcal{E}(||M^{\mathcal{B}}\tilde{R}||_{I},\xi)$,
and $\mu:=\mu\left(\mathcal{E},\xi\right)$. Consider the following
estimator design
\begin{align}
\mu= & \frac{\exp\left(2\mathcal{E}\right)+\exp\left(-2\mathcal{E}\right)+2}{8\xi\bar{\delta}}\label{eq:SO3_PPF_STCH_mu_VM}\\
\dot{\hat{R}}= & \hat{R}\left[\Omega_{m}-\hat{b}-W\right]_{\times},\quad\hat{R}\left(0\right)=\hat{R}_{0}\label{eq:SO3_PPF_STCH_Rest_dot_VM}\\
\dot{\hat{b}}= & \gamma_{1}\mu\left(\mathcal{E}+1\right)\exp\left(\mathcal{E}\right)\boldsymbol{\Upsilon}(M^{\mathcal{B}}\tilde{R})\label{eq:SO3_PPF_STCH_b_est_VM}\\
\dot{\hat{\sigma}}= & \gamma_{2}\left(\mathcal{E}+2\right)\exp\left(\mathcal{E}\right)\mu^{2}{\rm diag}\left(\boldsymbol{\Upsilon}(M^{\mathcal{B}}\tilde{R})\right)\boldsymbol{\Upsilon}(M^{\mathcal{B}}\tilde{R})\label{eq:SO3_PPF_STCH_s_est_VM}\\
W= & 2\frac{\mathcal{E}+2}{\mathcal{E}+1}\mu{\rm diag}\left(\boldsymbol{\Upsilon}(M^{\mathcal{B}}\tilde{R})\right)\hat{\sigma}\nonumber \\
& +\frac{4}{\underline{\lambda}}\frac{k_{w}\mu\mathcal{E}-\dot{\xi}/\xi}{1+\boldsymbol{\mathcal{J}}(M^{\mathcal{B}},\tilde{R})}\boldsymbol{\Upsilon}(M^{\mathcal{B}}\tilde{R})\label{eq:SO3_PPF_STCH_W_VM}
\end{align}
with $\boldsymbol{\mathcal{J}}(M^{\mathcal{B}},\tilde{R})$ and $\boldsymbol{\Upsilon}(M^{\mathcal{B}}\tilde{R})=\mathbf{vex}(\boldsymbol{\mathcal{P}}_{a}(M^{\mathcal{B}}\tilde{R}))$
being defined in \eqref{eq:SO3_PPF_STCH_Gamma} and \eqref{eq:SO3_PPF_STCH_VEX_VM},
respectively, $\xi$ being PPF given in \eqref{eq:SO3_PPF_STCH_Presc},
and $\underline{\lambda}:=\underline{\lambda}(\bar{\mathbf{M}}^{\mathcal{B}})$
being the minimum eigenvalue of $\bar{\mathbf{M}}^{\mathcal{B}}$.
$\mathcal{E}:=\mathcal{E}(||M^{\mathcal{B}}\tilde{R}||_{I},\xi)$
is defined in \eqref{eq:SO3_PPF_STCH_trans3}, while $\gamma_{1}$,
$\gamma_{2}$, and $k_{w}$ are positive constants. $\hat{b}$ and
$\hat{\sigma}$ are the estimates of $b$ and $\sigma$, respectively. The quaternion representation of the direct proposed observer is given in \nameref{sec:SO3_PPF_STCH_AppendixB}.

\begin{thm}
	\textbf{\label{thm:SO3_PPF_STCH_2}} Let the estimator in \eqref{eq:SO3_PPF_STCH_Rest_dot_VM},
	\eqref{eq:SO3_PPF_STCH_b_est_VM}, \eqref{eq:SO3_PPF_STCH_s_est_VM},
	and \eqref{eq:SO3_PPF_STCH_W_VM} be combined with the measurements
	in \eqref{eq:SO3_PPF_STCH_Vector_norm} and the angular velocity measurements
	in \eqref{eq:SO3_PPF_STCH_Angular}. Suppose that two or more instantaneous
	non-collinear body-frame vectors are available for measurements such
	that $M^{\mathcal{B}}$ has the rank of 3. Therefore, for $\tilde{R}\left(0\right)\notin\mathcal{S}$,
	as defined in Definition \ref{rem:Unstable-set}, and $\mathcal{E}\left(0\right)\in\mathbb{R}$,
	all error signals are bounded, while $\mathcal{E}\left(t\right)$
	asymptotically approaches $0$ and $\tilde{R}$ asymptotically approaches
	$\mathbf{I}_{3}$ in probability from almost any initial condition.
\end{thm}
In accordance with Theorem \ref{thm:SO3_PPF_STCH_2} the stability
of the estimator kinematics in \eqref{eq:SO3_PPF_STCH_Rest_dot_VM},
\eqref{eq:SO3_PPF_STCH_b_est_VM}, \eqref{eq:SO3_PPF_STCH_s_est_VM},
and \eqref{eq:SO3_PPF_STCH_W_VM} is guaranteed, since $\mathcal{E}\left(t\right)$
approaches the origin asymptotically. Consequently, $\mathcal{E}\left(t\right)$
is bounded and well-defined, which in turn implies that $||\tilde{R}||_{I}$
obeys the dynamic boundaries of transient and steady-state PPF as
defined in \eqref{eq:SO3_PPF_STCH_Presc} in consistence with Remark
\ref{rem:SO3_PPF_STCH_1}.

\textbf{Proof. }Let $\tilde{R}=R^{\top}\hat{R}$, $\tilde{b}=b-\hat{b}$,
and $\tilde{\sigma}=\sigma-\hat{\sigma}$ as defined in \eqref{eq:SO3_PPF_STCH_R_error},
\eqref{eq:SO3_PPF_STCH_b_tilde}, and \eqref{eq:SO3_PPF_STCH_s_tilde},
respectively. From \eqref{eq:SO3_PPF_STCH_R_dynam} and \eqref{eq:SO3_PPF_STCH_Rest_dot_VM},
the attitude error kinematics are analogous to \eqref{eq:SO3_PPF_STCH_Rtilde_dot}.
From \eqref{eq:SO3_PPF_STCH_MB_MI}, it can be found that
\begin{align}
\dot{M}^{\mathcal{B}} & =\dot{R}^{\top}M^{\mathcal{I}}R+R^{\top}M^{\mathcal{I}}\dot{R}\nonumber \\
& =-\left[\Omega\right]_{\times}R^{\top}M^{\mathcal{I}}R+R^{\top}M^{\mathcal{I}}R\left[\Omega\right]_{\times}\nonumber \\
& =-\left[\Omega\right]_{\times}M^{\mathcal{B}}+M^{\mathcal{B}}\left[\Omega\right]_{\times}\label{eq:SO3_PPF_STCH_MB_dot}
\end{align}
Hence, from \eqref{eq:SO3_PPF_STCH_Rtilde_dot} and \eqref{eq:SO3_PPF_STCH_MB_dot},
one has
\begin{align}
\frac{d}{dt}||M^{\mathcal{B}}\tilde{R}||_{I}= & -\frac{1}{4}{\rm Tr}\{\dot{M}^{\mathcal{B}}\tilde{R}+M^{\mathcal{B}}\dot{\tilde{R}}\}\label{eq:SO3_PPF_STCH_NormMBRtilde_dot}
\end{align}
such that
\begin{align}
d||M^{\mathcal{B}}\tilde{R}||_{I}= & -\frac{1}{4}{\rm Tr}\left\{ \left(-\left[\Omega\right]_{\times}M^{\mathcal{B}}+M^{\mathcal{B}}\left[\Omega\right]_{\times}\right)\tilde{R}\right\} dt\nonumber \\
& -\frac{1}{4}{\rm Tr}\left\{ M^{\mathcal{B}}\left(\tilde{R}\left[\Omega\right]_{\times}+\left[\Omega\right]_{\times}^{\top}\tilde{R}\right)\right\} dt\nonumber \\
& -\frac{1}{4}{\rm Tr}\left\{ M^{\mathcal{B}}\tilde{R}\left[\tilde{b}-W\right]_{\times}dt+M^{\mathcal{B}}\tilde{R}\left[\mathcal{Q}d\beta\right]_{\times}\right\} \nonumber \\
= & -\frac{1}{4}{\rm Tr}\left\{ M^{\mathcal{B}}\tilde{R}\left[\tilde{b}-W\right]_{\times}dt+M^{\mathcal{B}}\tilde{R}\left[\mathcal{Q}d\beta\right]_{\times}\right\} \nonumber \\
& -\frac{1}{4}{\rm Tr}\left\{ \left[M^{\mathcal{B}}\tilde{R},\left[\Omega\right]_{\times}\right]\right\} dt\nonumber \\
= & \frac{1}{2}\boldsymbol{\Upsilon}(M^{\mathcal{B}}\tilde{R})^{\top}\left((\tilde{b}-W)dt+\mathcal{Q}d\beta\right)\label{eq:SO3_PPF_STCH_NormMBRtilde_dot-1}
\end{align}
Recalling the identities in  \eqref{eq:SO3_PPF_STCH_Identity7} and
\eqref{eq:SO3_PPF_STCH_Identity5}, it becomes apparent that ${\rm Tr}\{M^{\mathcal{B}}\tilde{R}[\tilde{b}]_{\times}\}=-2\boldsymbol{\Upsilon}(M^{\mathcal{B}}\tilde{R})^{\top}\tilde{b}$
and ${\rm Tr}\left\{ [M^{\mathcal{B}}\tilde{R},\left[\Omega\right]_{\times}]\right\} =0$,
respectively. Thereby, in view of \eqref{eq:SO3_PPF_STCH_NormR_dynam}
and \eqref{eq:SO3_PPF_STCH_Trans_dot}, and considering the result
in \eqref{eq:SO3_PPF_STCH_NormMBRtilde_dot-1}, the transformed error
can be expressed in the incremental form as follows: 
\begin{align}
d\mathcal{E} & =f(\mathcal{E},\tilde{b})dt+g\left(\mathcal{E}\right)\mathcal{Q}d\beta\label{eq:SO3_PPF_STCH_Trans_dot_VM_Ito}
\end{align}
where $f(\mathcal{E},\tilde{b})=2\mu\left(\frac{1}{2}\boldsymbol{\Upsilon}(M^{\mathcal{B}}\tilde{R})^{\top}(\tilde{b}-W)-\frac{\dot{\xi}}{\xi}||M^{\mathcal{B}}\tilde{R}||_{I}\right)$
and $g\left(\mathcal{E}\right)=\mu\boldsymbol{\Upsilon}(M^{\mathcal{B}}\tilde{R})^{\top}$.
Assuming that the dynamics in \eqref{eq:SO3_PPF_STCH_Trans_dot_VM_Ito}
are presented in the sense of Stratonovich \cite{stratonovich1967topics},
its transformation to Ito \cite{ito1984lectures} can be expressed
as follows
\begin{align}
d\mathcal{E} & =\mathcal{F}(\mathcal{E},\tilde{b})dt+g\left(\mathcal{E}\right)\mathcal{Q}d\beta\label{eq:SO3_PPF_STCH_Trans_dot_VM_Strat}
\end{align}
such that
\[
\mathcal{F}(\mathcal{E},\tilde{b})=f(\mathcal{E},\tilde{b})+\boldsymbol{\mathcal{W}}\left(\mathcal{E}\right)
\]
with $\boldsymbol{\mathcal{W}}\left(\mathcal{E}\right)$ being the
Wong-Zakai factor. In view of \eqref{eq:SO3_PPF_STCH_Wong_STRAT},
$\boldsymbol{\mathcal{W}}\left(\mathcal{E}\right)$ is defined in
the following manner
\[
\boldsymbol{\mathcal{W}}\left(\mathcal{E}\right)=\frac{\exp\left(2\mathcal{E}\right)-\exp\left(-2\mathcal{E}\right)}{8\xi\bar{\delta}}\mu\boldsymbol{\Upsilon}(M^{\mathcal{B}}\tilde{R})^{\top}\mathcal{Q}^{2}\boldsymbol{\Upsilon}(M^{\mathcal{B}}\tilde{R})
\]
Consider the following candidate Lyapunov function

\begin{align}
V(\mathcal{E},\tilde{b},\tilde{\sigma}) & =\mathcal{E}\exp\left(\mathcal{E}\right)+\frac{1}{2\gamma_{1}}||\tilde{b}||^{2}+\frac{1}{\gamma_{2}}||\tilde{\sigma}||^{2}\label{eq:SO3_PPF_STCH_V_VM}
\end{align}
For $V:=V(\mathcal{E},\tilde{b},\tilde{\sigma})$, the first and second
partial derivatives of \eqref{eq:SO3_PPF_STCH_V_VM} are equivalent
to \eqref{eq:SO3_PPF_STCH_Vv_Ry} and \eqref{eq:SO3_PPF_STCH_Vvv_Ry},
respectively. Thus, from \eqref{eq:SO3_PPF_STCH_V_VM}, \eqref{eq:SO3_PPF_STCH_Trans_dot_VM_Strat},
\eqref{eq:SO3_PPF_STCH_Vv_Ry}, and \eqref{eq:SO3_PPF_STCH_Vvv_Ry},
the differential operator $\mathcal{L}V$ becomes
\begin{align}
\mathcal{L}V= & V_{\mathcal{E}}^{\top}\mathcal{F}(\mathcal{E},\tilde{b})+\frac{1}{2}{\rm Tr}\{g\left(\mathcal{E}\right)\mathcal{Q}^{2}g^{\top}\left(\mathcal{E}\right)V_{\mathcal{E}\mathcal{E}}\}\nonumber \\
& -\frac{1}{\gamma_{1}}\tilde{b}^{\top}\dot{\hat{b}}-\frac{2}{\gamma_{2}}\tilde{\sigma}^{\top}\dot{\hat{\sigma}}\label{eq:SO3_PPF_STCH_Vdot_VM_0}
\end{align}
which is
\begin{align}
\mathcal{L}V= & \left(\mathcal{E}+1\right)\exp\left(\mathcal{E}\right)\mu\left(\boldsymbol{\Upsilon}(M^{\mathcal{B}}\tilde{R})^{\top}(\tilde{b}-W)\right.\nonumber \\
& \left.+\frac{\exp\left(2\mathcal{E}\right)-\exp\left(-2\mathcal{E}\right)}{8\xi\bar{\delta}}\boldsymbol{\Upsilon}(M^{\mathcal{B}}\tilde{R})^{\top}\mathcal{Q}^{2}\boldsymbol{\Upsilon}(M^{\mathcal{B}}\tilde{R})\right)\nonumber \\
& -2\left(\mathcal{E}+1\right)\exp\left(\mathcal{E}\right)\mu\frac{\dot{\xi}}{\xi}||M^{\mathcal{B}}\tilde{R}||_{I}\nonumber \\
& +\frac{1}{2}\left(\mathcal{E}+2\right)\exp\left(\mathcal{E}\right)\mu^{2}\boldsymbol{\Upsilon}(M^{\mathcal{B}}\tilde{R})^{\top}\mathcal{Q}^{2}\boldsymbol{\Upsilon}(M^{\mathcal{B}}\tilde{R})\nonumber \\
& -\frac{1}{\gamma_{1}}\tilde{b}^{\top}\dot{\hat{b}}-\frac{2}{\gamma_{2}}\tilde{\sigma}^{\top}\dot{\hat{\sigma}}\label{eq:SO3_PPF_STCH_Vdot_VM_1}
\end{align}
Recall that, $\mathcal{E}>0\forall||M^{\mathcal{B}}\tilde{R}||_{I}\neq0$
and $\mathcal{E}=0$ only at $||M^{\mathcal{B}}\tilde{R}||_{I}=0$,
according to property (ii) and (iii) of Proposition \ref{Prop:SO3_PPF_STCH_1}.
In this regard, from \eqref{eq:SO3_PPF_STCH_mu} and \eqref{eq:SO3_PPF_STCH_mu_VM}
one has $\mu>0\forall t\geq0$. This implies that $\mathcal{E}\exp\left(\mathcal{E}\right)\geq0$,
$\mu>0$ and $\boldsymbol{\Upsilon}(M^{\mathcal{B}}\tilde{R})^{\top}\mathcal{Q}^{2}\boldsymbol{\Upsilon}(M^{\mathcal{B}}\tilde{R})\geq0$
for all $t\geq0$. In addition, it can be deduced that $\mu>\frac{\exp\left(2\mathcal{E}\right)-\exp\left(-2\mathcal{E}\right)}{8\xi\bar{\delta}}$.
Accordingly, the differential operator in \eqref{eq:SO3_PPF_STCH_Vdot_VM_1}
could be expressed in a form of inequality as follows
\begin{align}
\mathcal{L}V\leq & \left(\mathcal{E}+1\right)\exp\left(\mathcal{E}\right)\mu\boldsymbol{\Upsilon}(M^{\mathcal{B}}\tilde{R})^{\top}(\tilde{b}-W)\nonumber \\
& -2\left(\mathcal{E}+1\right)\exp\left(\mathcal{E}\right)\mu\frac{\dot{\xi}}{\xi}||M^{\mathcal{B}}\tilde{R}||_{I}\nonumber \\
& +2\left(\mathcal{E}+2\right)\exp\left(\mathcal{E}\right)\mu^{2}\boldsymbol{\Upsilon}(M^{\mathcal{B}}\tilde{R})^{\top}{\rm diag}\left(\boldsymbol{\Upsilon}(M^{\mathcal{B}}\tilde{R})\right)\sigma\nonumber \\
& -\frac{1}{\gamma_{1}}\tilde{b}^{\top}\dot{\hat{b}}-\frac{2}{\gamma_{2}}\tilde{\sigma}^{\top}\dot{\hat{\sigma}}\label{eq:SO3_PPF_STCH_Vdot_VM_2}
\end{align}
with $\sigma$ being defined in \eqref{eq:SO3_PPF_STCH_s_factor}.
Directly substituting $\dot{\hat{b}}$, $\dot{\hat{\sigma}}$, and
$W$ with their definitions in \eqref{eq:SO3_PPF_STCH_b_est_VM},
\eqref{eq:SO3_PPF_STCH_s_est_VM}, and \eqref{eq:SO3_PPF_STCH_W_VM},
respectively, transforms the inequality above as follows
\begin{align}
\mathcal{L}V\leq & 2\left(\mathcal{E}+1\right)\exp\left(\mathcal{E}\right)\mu\frac{\dot{\xi}}{\xi}\left(\frac{2}{\underline{\lambda}}\frac{\left\Vert \boldsymbol{\Upsilon}(M^{\mathcal{B}}\tilde{R})\right\Vert ^{2}}{1+\boldsymbol{\mathcal{J}}(M^{\mathcal{B}},\tilde{R})}-||M^{\mathcal{B}}\tilde{R}||_{I}\right)\nonumber \\
& -\frac{2}{\underline{\lambda}}k_{w}\exp\left(\mathcal{E}\right)\left(\mathcal{E}+1\right)^{2}\mu^{2}\frac{\left\Vert \boldsymbol{\Upsilon}(M^{\mathcal{B}}\tilde{R})\right\Vert ^{2}}{1+\boldsymbol{\mathcal{J}}(M^{\mathcal{B}},\tilde{R})}\label{eq:SO3_PPF_STCH_Vdot_VM_3}
\end{align}
It can be easily found that
\begin{equation}
\left(\mathcal{E}+1\right)\exp\left(\mathcal{E}\right)\mu\frac{\dot{\xi}}{\xi}\left(\frac{2}{\underline{\lambda}}\frac{\left\Vert \boldsymbol{\Upsilon}(M^{\mathcal{B}}\tilde{R})\right\Vert ^{2}}{1+\boldsymbol{\mathcal{J}}(M^{\mathcal{B}},\tilde{R})}-||M^{\mathcal{B}}\tilde{R}||_{I}\right)\leq0\label{eq:SO3_PPF_STCH_Factor}
\end{equation}
where $\dot{\xi}$ is a negative strictly increasing component which
satisfies $\dot{\xi}\rightarrow0$ as $t\rightarrow\infty$, while
$\xi:\mathbb{R}_{+}\to\mathbb{R}_{+}$ such that $\xi\rightarrow\xi_{\infty}$
as $t\rightarrow\infty$. Thus, $\dot{\xi}/\xi\leq0$. According to
the equation \eqref{eq:SO3_PPF_STCH_lemm1_3} of Lemma \ref{Lemm:SO3_PPF_STCH_1},
the expression in \eqref{eq:SO3_PPF_STCH_Factor} is negative semi-definite.
Thus, the inequality in \eqref{eq:SO3_PPF_STCH_Vdot_VM_3} becomes
\begin{align}
\mathcal{L}V\leq & -k_{w}\exp\left(\mathcal{E}\right)\left(\mathcal{E}+1\right)\mathcal{E}\mu^{2}||M^{\mathcal{B}}\tilde{R}||_{I}\label{eq:SO3_PPF_STCH_Vdot_VM_4}
\end{align}
Due to the fact that 
\begin{equation}
||M^{\mathcal{B}}\tilde{R}||_{I}=\xi\frac{\bar{\delta}\exp\left(\mathcal{E}\right)-\underline{\delta}\exp\left(-\mathcal{E}\right)}{\exp\left(\mathcal{E}\right)+\exp\left(-\mathcal{E}\right)}\label{eq:SO3_PPF_STCH_e_VM}
\end{equation}
The inequality in \eqref{eq:SO3_PPF_STCH_Vdot_VM_4} can be expressed
as
\begin{align}
\mathcal{L}V\leq & -\bar{\delta}k_{w}\xi\mu^{2}\left(\mathcal{E}+1\right)\mathcal{E}\frac{\exp\left(\mathcal{E}\right)-\exp\left(-\mathcal{E}\right)}{\exp\left(\mathcal{E}\right)+\exp\left(-\mathcal{E}\right)}\exp\left(\mathcal{E}\right)\label{eq:SO3_PPF_STCH_Vdot_VM_Final}
\end{align}
where $\bar{\delta}>0$, $k_{w}>0$, and $\xi:\mathbb{R}_{+}\to\mathbb{R}_{+}$.
On the basis of property (ii) and (iii) of Proposition \ref{Prop:SO3_PPF_STCH_1},
$\mathcal{E}>0\forall||M^{\mathcal{B}}\tilde{R}||_{I}\neq0$ and $\mathcal{E}=0$
only at $||M^{\mathcal{B}}\tilde{R}||_{I}=0$ for $\tilde{R}\left(0\right)\notin\mathcal{S}$
and $\mathcal{E}\left(0\right)\in\mathbb{R}$. Thus, $\mathcal{E}=0$
implies that $\tilde{R}=\mathbf{I}_{3}$. Since $\mathcal{L}V$ is
bounded and $V$ is radially unbounded, it can be stated that for
$\tilde{R}\left(0\right)\notin\mathcal{S}$ and $\mathcal{E}\left(0\right)\in\mathbb{R}$
there exists a unique strong solution to the stochastic system in
\eqref{eq:SO3_PPF_STCH_Trans_dot_VM_Strat} with the probability of
one \cite{khasminskii1980stochastic}. Thus, from the result in \eqref{eq:SO3_PPF_STCH_Vdot_VM_Final},
$\mathcal{E}=0$ as well as $\tilde{R}=\mathbf{I}_{3}$ are independent
of the unknown noise and bias attached to angular velocity measurements.
Thereby, based on the stochastic LaSalle Theorem \cite{khasminskii1980stochastic},
one has $\mathbb{P}\{\lim_{t\rightarrow\infty}\left|\mathcal{E}\left(t\right)\right|=0\}=1,\forall\tilde{R}\left(0\right)\notin\mathcal{S},\mathcal{E}\left(0\right)\in\mathbb{R}$
which in turn indicates that $\mathbb{P}\{\lim_{t\rightarrow\infty}\tilde{R}=\mathbf{I}_{3}\}=1$
for all $\tilde{R}\left(0\right)\notin\mathcal{S}$ and $\mathcal{E}\left(0\right)\in\mathbb{R}$.
Furthermore, the estimation of $b$ and $\sigma$ has a finite limit
in probability (\cite{deng2001stabilization}, Theorem 3.1).

\subsection{Remarks on Stability Analysis and Implementation Steps}

It becomes apparent that the gains associated with the vex operator
$\boldsymbol{\Upsilon}(\cdot)=\mathbf{vex}(\boldsymbol{\mathcal{P}}_{a}(\cdot))$
of $\dot{\hat{b}}$, $\dot{\hat{\sigma}}$, and $W$ in \eqref{eq:SO3_PPF_STCH_b_est_Ry},
\eqref{eq:SO3_PPF_STCH_s_est_Ry}, and \eqref{eq:SO3_PPF_STCH_W_Ry},
or in \eqref{eq:SO3_PPF_STCH_b_est_VM}, \eqref{eq:SO3_PPF_STCH_s_est_VM},
and \eqref{eq:SO3_PPF_STCH_W_VM}, respectively, are dynamic. Their
dynamic behavior forces the attitude error to comply with the prescribed
performance constraints. Consequently, both proposed estimators are
characterized by highly favorable features which cause the dynamic
gains to become increasingly aggressive as $||\tilde{R}||_{I}$ approaches
the unstable equilibria +1. On the other side, these gains decrease
significantly as $||\tilde{R}||_{I}\rightarrow0$. This dynamic behavior
allows the proposed nonlinear stochastic estimators to force the attitude
error to follow the predefined PPF imposed by the user and thereby
to achieve the predetermined measures of transient as well as steady-state
performance. Let us summarize the design process of the estimator
proposed in Subsection \ref{subsec:SO3_PPF_STCH_Passive-Filter}:
\begin{enumerate}
	\item[\textbf{1.}] Set the following parameters: $\bar{\delta}=\underline{\delta}>||\tilde{R}\left(0\right)||_{I}$,
	$\xi_{0}=\bar{\delta}$, a small set $\xi_{\infty}$ and a convergence
	rate $\ell$.  
	\item[\textbf{2.}] Evaluate $\boldsymbol{\Upsilon}(\tilde{R})=\mathbf{vex}(\boldsymbol{\mathcal{P}}_{a}(\tilde{R}))$
	and $||\tilde{R}||_{I}=\frac{1}{4}{\rm Tr}\{\mathbf{I}_{3}-\tilde{R}\}$.
	
	\item[\textbf{3.}] Evaluate the PPF $\xi$ from \eqref{eq:SO3_PPF_STCH_Presc}.  
	\item[\textbf{4.}] Evaluate $\mu(||\tilde{R}||_{I},\xi)$ and $\mathcal{E}(||\tilde{R}||_{I},\xi)$
	from \eqref{eq:SO3_PPF_STCH_mu1} and \eqref{eq:SO3_PPF_STCH_trans3},
	respectively.  
	\item[\textbf{5.}] Evaluate the estimator design $\dot{\hat{R}}$, $\dot{\hat{b}}$,
	$\dot{\hat{\sigma}}$ and $W$ from \eqref{eq:SO3_PPF_STCH_Rest_dot_Ry},
	\eqref{eq:SO3_PPF_STCH_b_est_Ry}, \eqref{eq:SO3_PPF_STCH_s_est_Ry},
	and \eqref{eq:SO3_PPF_STCH_W_Ry}, respectively.  
	\item[\textbf{6.}] Go to \textbf{Step 2}.
\end{enumerate}
In the same manner, the steps stated above are applicable for the
estimator proposed in Subsection \ref{subsec:SO3_PPF_STCH_Explicit-Filter}.

\subsection{Stochastic Estimators with Prescribed Performance: Discrete-form\label{Subsec:SO3PPF-Discrete}}

For the purpose of implementation, this subsection presents a discrete-form
of the stochastic estimators proposed in Subsection \ref{subsec:SO3_PPF_STCH_Passive-Filter}
and Subsection \ref{subsec:SO3_PPF_STCH_Explicit-Filter}. The exact
integration of \eqref{eq:SO3_PPF_STCH_R_dynam} is equivalent to
\begin{equation}
R\left[k+1\right]=R\left[k\right]\exp\left(\left[\Omega\left[k\right]\right]_{\times}\Delta t\right)\label{eq:Comparison_R_discrete}
\end{equation}
where $\Delta t$ is a small time sample and $\left[k\right]$ associated
with a variable refers to its value at the $k$th sample for $k\in\mathbb{N}$.
In view of \eqref{eq:Comparison_R_discrete}, the discrete analogue
of the semi-direct nonlinear stochastic estimator with prescribed
performance outlined in \eqref{eq:SO3_PPF_STCH_Rest_dot_VM}, \eqref{eq:SO3_PPF_STCH_b_est_VM},
\eqref{eq:SO3_PPF_STCH_s_est_VM}, and \eqref{eq:SO3_PPF_STCH_W_VM},
is given in Algorithm \ref{alg:Discrete-nonlinear-filter}.

\begin{algorithm}
	\caption{\label{alg:Discrete-nonlinear-filter}Complete implementation steps
		of the proposed direct stochastic estimator}
	
	\textbf{Initialization}:
	\begin{enumerate}
		\item[{\footnotesize{}1:}] Set $\hat{R}[0]\in\mathbb{SO}\left(3\right)$. Alternative solution,
		construct $\hat{R}[0]\in\mathbb{SO}\left(3\right)$ using one of the methods
		of attitude determination, visit \cite{hashim2020AtiitudeSurvey}\vspace{1mm}
		\item[{\footnotesize{}2:}] Set $\hat{b}[0]=0_{3\times1}$ and $\hat{\sigma}[0]=0_{3\times1}$ \vspace{1mm}
		\item[{\footnotesize{}3:}] Select $\xi_{0}=\bar{\delta}=\underline{\delta}$, $\xi_{\infty}$,
		$\ell$, $k_{w}$, $\gamma_{1}$, and $\gamma_{2}$ as positive constants.
		Also, $s_{i}\geq0$ with $\sum_{i=1}^{n}s_{i}=3$.
	\end{enumerate}
	\textbf{while$\hspace{6em}$$\forall i=1,2,\ldots,n$}
	\begin{enumerate}
		\item[{\footnotesize{}4:}] Consider the measurements and observations in \eqref{eq:SO3_PPF_STCH_Vect_True}
		and compute the normalization $\upsilon_{i}^{\mathcal{I}}=\frac{{\rm v}_{i}^{\mathcal{I}}}{||{\rm v}_{i}^{\mathcal{I}}||}$
		and $\upsilon_{i}^{\mathcal{B}}=\frac{{\rm v}_{i}^{\mathcal{B}}}{||{\rm v}_{i}^{\mathcal{B}}||}$
		as in \eqref{eq:SO3_PPF_STCH_Vector_norm}\vspace{1mm}
		\item[{\footnotesize{}5:}] $\hat{\upsilon}_{i}^{\mathcal{B}}=\hat{R}^{\top}\upsilon_{i}^{\mathcal{I}}$
		as in \eqref{eq:SO3_PPF_STCH_vB_hat}\vspace{1mm}
		\item[{\footnotesize{}6:}] $M^{\mathcal{B}}=\sum_{i=1}^{n}s_{i}\upsilon_{i}^{\mathcal{B}}\left(\upsilon_{i}^{\mathcal{B}}\right)^{\top}$
		as in \eqref{eq:SO3_PPF_STCH_MB_MI} with
		\item[] $\bar{\mathbf{M}}^{\mathcal{B}}={\rm Tr}\{M^{\mathcal{B}}\}\mathbf{I}_{3}-M^{\mathcal{B}}$\vspace{1mm}
		\item[{\footnotesize{}7:}] $\boldsymbol{\Upsilon}=\sum_{i=1}^{n}\frac{s_{i}}{2}\hat{\upsilon}_{i}^{\mathcal{B}}\times\upsilon_{i}^{\mathcal{B}}$
		as in \eqref{eq:SO3_PPF_STCH_VEX_VM}\vspace{1mm}
		\item[{\footnotesize{}8:}] {\scriptsize{}$\boldsymbol{\mathcal{J}}={\rm Tr}\left\{ \left(\sum_{i=1}^{n}s_{i}\upsilon_{i}^{\mathcal{B}}\left(\upsilon_{i}^{\mathcal{B}}\right)^{\top}\right)^{-1}\sum_{i=1}^{n}s_{i}\upsilon_{i}^{\mathcal{B}}\left(\hat{\upsilon}_{i}^{\mathcal{B}}\right)^{\top}\right\} $}{\footnotesize{}
		}as in \eqref{eq:SO3_PPF_STCH_Gamma}\vspace{1mm}
		\item[{\footnotesize{}9:}] $||M^{\mathcal{B}}\tilde{R}||_{I}=\sum_{i=1}^{n}s_{i}\left(1-\left(\hat{\upsilon}_{i}^{\mathcal{B}}\right)^{\top}\upsilon_{i}^{\mathcal{B}}\right)$
		as in \eqref{eq:SO3_PPF_STCH_Gamma}\vspace{1mm}
		\item[{\footnotesize{}10:}] $\xi=\left(\xi_{0}-\xi_{\infty}\right)\exp\left(-\ell k\Delta t\right)+\xi_{\infty}$
		as in \eqref{eq:SO3_PPF_STCH_Presc} and
		\item[] $\xi_{d}=(\xi[k]-\xi[k-1])/\Delta t$\vspace{1mm}
		\item[{\footnotesize{}11:}] $\mathcal{E}[k]=\frac{1}{2}\text{ln}\frac{\underline{\delta}+||M^{\mathcal{B}}\tilde{R}||_{I}/\xi}{\bar{\delta}-||M^{\mathcal{B}}\tilde{R}||_{I}/\xi}$
		as in \eqref{eq:SO3_PPF_STCH_trans3}\vspace{1mm}
		\item[{\footnotesize{}12:}] $\mu=\frac{\exp\left(2\mathcal{E}[k]\right)+\exp\left(-2\mathcal{E}[k]\right)+2}{8\xi\bar{\delta}}$
		as in \eqref{eq:SO3_PPF_STCH_mu}\vspace{1mm}
		\item[{\footnotesize{}13:}] $W[k]=2\frac{\mathcal{E}[k]+2}{\mathcal{E}[k]+1}\mu{\rm diag}\left(\boldsymbol{\Upsilon}\right)\hat{\sigma}[k]+\frac{4}{\underline{\lambda}}\frac{k_{w}\mu\mathcal{E}[k]-\xi_{d}/\xi}{1+\boldsymbol{\mathcal{J}}}\boldsymbol{\Upsilon}$\vspace{1mm}
		\item[{\footnotesize{}14:}] $\hat{\Omega}[k]=\Omega_{m}[k]-\hat{b}[k]-W[k]$\vspace{1mm}
		\item[{\footnotesize{}15:}] $\hat{R}[k+1]=\hat{R}[k]\exp\left([\hat{\Omega}[k]]_{\times}\Delta t\right)$,
		visit \eqref{eq:SLAM_Compute_exp}\vspace{1mm}
		\item[{\footnotesize{}16:}] $\hat{b}[k+1]=\hat{b}_{\Omega}[k]+\Delta t\gamma_{1}\mu\left(\mathcal{E}[k]+1\right)\exp\left(\mathcal{E}[k]\right)\boldsymbol{\Upsilon}$\vspace{1mm}
		\item[{\footnotesize{}17:}] $\hat{\sigma}[k+1]=\hat{\sigma}[k]$
		\item[] $\hspace{4em}+\Delta t\gamma_{2}\left(\mathcal{E}[k]+2\right)\exp\left(\mathcal{E}[k]\right)\mu^{2}{\rm diag}\left(\boldsymbol{\Upsilon}\right)\boldsymbol{\Upsilon}$
		\item[{\footnotesize{}18:}] $k+1\rightarrow k$
	\end{enumerate}
	\textbf{end while}
\end{algorithm}
Define $x\in\mathbb{R}^{3}$ as an axis that is rotating by an angle
$\beta\in\mathbb{R}$ in a 2-sphere $\mathbb{S}^{2}$. The rigid-body's
attitude is given by $\mathcal{R}_{\beta}:\mathbb{R}\times\mathbb{R}^{3}\rightarrow\mathbb{SO}\left(3\right)$
visit \eqref{eq:SO3_PPF_STCH_att_ang}
\begin{align*}
\mathcal{R}_{\beta}\left(\beta,x\right) & =\mathbf{I}_{3}+\sin\left(\beta\right)\left[x\right]_{\times}+\left(1-\cos\left(\beta\right)\right)\left[x\right]_{\times}^{2}
\end{align*}
The expression $\hat{R}[k+1]=\hat{R}[k]\exp([\hat{\Omega}[k]]_{\times}\Delta t)$
uses exact integration. Thereby, to guarantee that $\exp([\hat{\Omega}[k]]_{\times}\Delta t)\in\mathbb{SO}\left(3\right)$,
define $u=\hat{\Omega}[k]\Delta t$, where $x=u/\left\Vert u\right\Vert $
and $\beta=\left\Vert u\right\Vert $. Hence, $\exp([\hat{\Omega}[k]]_{\times}\Delta t)=\mathcal{R}_{\beta}\left(\beta,x\right)$
can be obtained as follows
\begin{align}
\exp\left([\hat{\Omega}[k]]_{\times}\Delta t\right) & =\mathbf{I}_{3}+\sin\left(\beta\right)\left[x\right]_{\times}+\left(1-\cos\left(\beta\right)\right)\left[x\right]_{\times}^{2}\label{eq:SLAM_Compute_exp}
\end{align}
Likewise, in the light of Algorithm, the discrete filter of \eqref{eq:SO3_PPF_STCH_mu_Ry},
\eqref{eq:SO3_PPF_STCH_Rest_dot_Ry}, \eqref{eq:SO3_PPF_STCH_b_est_Ry},
\eqref{eq:SO3_PPF_STCH_s_est_Ry}, and \eqref{eq:SO3_PPF_STCH_W_Ry}
is given by
\begin{equation}
\begin{cases}
\mu\left[k\right] & =\frac{\exp\left(2\mathcal{E}\left[k\right]\right)+\exp\left(-2\mathcal{E}\left[k\right]\right)+2}{8\xi\left[k\right]\bar{\delta}}\\
\hat{R}\left[k+1\right] & =\hat{R}\left[k\right]\exp\left(\left[\Omega_{m}\left[k\right]-\hat{b}\left[k\right]-W\left[k\right]\right]_{\times}\Delta t\right)\\
W\left[k\right] & =2\frac{\mathcal{E}\left[k\right]+2}{\mathcal{E}\left[k\right]+1}\mu\left[k\right]{\rm diag}\left(\boldsymbol{\Upsilon}(\left[k\right])\right)\hat{\sigma}\left[k\right]\\
& \hspace{1em}+2\frac{k_{w}\mu\left[k\right]\left(\mathcal{E}\left[k\right]+1\right)-\bar{\xi}_{d}\left[k\right]/4\xi\left[k\right]}{1-||\tilde{R}\left[k\right]||_{I}}\boldsymbol{\Upsilon}(\tilde{R}\left[k\right])\\
\hat{b}\left[k+1\right] & =\hat{b}\left[k\right]\\
& \hspace{1em}+\gamma_{1}\left(\mathcal{E}\left[k\right]+1\right)\exp\left(\mathcal{E}\left[k\right]\right)\mu\left[k\right]\boldsymbol{\Upsilon}(\tilde{R}\left[k\right])\Delta t\\
\hat{\sigma}\left[k+1\right] & =\hat{\sigma}\left[k\right]+\gamma_{2}\mathcal{E}\left[k\right]\left(\mathcal{E}\left[k\right]+2\right)\exp\left(\mathcal{E}\left[k\right]\right)\mu^{2}\left[k\right]\\
& \hspace{6em}\times{\rm diag}\left(\boldsymbol{\Upsilon}(\tilde{R}\left[k\right])\right)\boldsymbol{\Upsilon}(\tilde{R}\left[k\right])\Delta t
\end{cases}\label{eq:Comp_Non_GPSd_NDAF-1-1}
\end{equation}
with 
\begin{equation}
\begin{cases}
\xi\left[k\right] & =\left(\xi_{0}-\xi_{\infty}\right)\exp\left(-\ell k\right)+\xi_{\infty}\\
\bar{\xi}_{d}\left[k\right] & =\frac{\xi\left[k\right]-\xi\left[k-1\right]}{\Delta t}\\
\mathcal{E}\left[k\right] & =\frac{1}{2}\text{ln}\frac{\underline{\delta}+||\tilde{R}\left[k\right]||_{I}/\xi\left[k\right]}{\bar{\delta}-||\tilde{R}\left[k\right]||_{I}/\xi\left[k\right]}
\end{cases}\label{eq:Comp_Non_PPF_NDAF-1-1-1}
\end{equation}

\section{Simulations \label{sec:SO3_PPF_STCH_Simulations}}

In this section the performance of the proposed nonlinear stochastic
attitude estimators on $\mathbb{SO}\left(3\right)$ with prescribed
performance is examined and tested against large initialization error
and high level of noise and bias components attached to the measurements.
Let the true attitude dynamics be as in \eqref{eq:SO3_PPF_STCH_R_dynam}
with the following angular velocity 
\[
\Omega=\left[{\rm sin}\left(0.4t\right),{\rm sin}\left(0.7t+\frac{\pi}{4}\right),0.4{\rm cos}\left(0.3t\right)\right]^{\top}\left({\rm rad/sec}\right)
\]
where $R\left(0\right)=\mathbf{I}_{3}$. Let the measurement of angular
velocity be $\Omega_{m}=\Omega+b+\omega$ with $b=0.1\left[1,-1,1\right]^{\top}$
and $\omega$ being a wide-band of random white noise process with
a standard deviation (STD) $0.3\left({\rm rad/sec}\right)$. Consider
two non-collinear inertial-frame vectors given by ${\rm v}_{1}^{\mathcal{I}}=\frac{1}{\sqrt{3}}\left[1,-1,1\right]^{\top}$
and ${\rm v}_{2}^{\mathcal{I}}=\left[0,0,1\right]^{\top}$, while
their measured values in the body-frame are ${\rm v}_{i}^{\mathcal{B}}=R^{\top}{\rm v}_{i}^{\mathcal{I}}+{\rm b}_{i}^{\mathcal{B}}+\omega_{i}^{\mathcal{B}}$
for $i=1,2$. Let ${\rm b}_{1}^{\mathcal{B}}=0.1\left[-1,1,0.5\right]^{\top}$
and ${\rm b}_{2}^{\mathcal{B}}=0.1\left[0,0,1\right]^{\top}$and suppose
that $\omega_{i}^{\mathcal{B}}$ is a white noise vector with zero
mean and a ${\rm STD=}0.12$ for $i=1,2$. Let $\upsilon_{i}^{\mathcal{I}}={\rm v}_{i}^{\mathcal{I}}/||{\rm v}_{i}^{\mathcal{I}}||$
and $\upsilon_{i}^{\mathcal{B}}={\rm v}_{i}^{\mathcal{B}}/||{\rm v}_{i}^{\mathcal{B}}||$
for $i=1,2$ with $\upsilon_{3}^{\mathcal{I}}=\upsilon_{1}^{\mathcal{I}}\times\upsilon_{2}^{\mathcal{I}}$
and $\upsilon_{3}^{\mathcal{B}}=\upsilon_{1}^{\mathcal{B}}\times\upsilon_{2}^{\mathcal{B}}$.
Also, $s_{1}=1.4$, $s_{2}=1.4$, and $s_{3}=0.2$. The given measurements
are uncertain and signify gyro and two vector measurements which is
characteristic of a low-cost IMU module. For the semi-direct estimator,
the reconstructed attitude $R_{y}$ is evaluated by SVD \cite{markley1988attitude}
such that $\tilde{R}=R_{y}^{\top}\hat{R}$. The total simulation time
is 30 seconds.

A very large initial attitude error has been considered with the initial
attitude rotation of $\hat{R}$ defined according to angle-axis parameterization
in \eqref{eq:SO3_PPF_STCH_att_ang}. $\hat{R}\left(0\right)=\mathcal{R}_{\theta}(\theta,u/||u||)$
where $\theta=178({\rm deg})$ and $u=\left[4,1,5\right]^{\top}$,
and $||\tilde{R}||_{I}$ is very near to the unstable equilibria (+1)
with $||\tilde{R}||_{I}\approx0.9999$
\[
R\left(0\right)=\mathbf{I}_{3},\hspace{1em}\hat{R}\left(0\right)=\left[\begin{array}{ccc}
-0.2377 & 0.1635 & 0.9575\\
0.2173 & -0.9518 & 0.2165\\
0.9467 & 0.2596 & 0.1907
\end{array}\right]
\]
Initial estimates are $\hat{b}\left(0\right)=\left[0,0,0\right]^{\top}$
and $\hat{\sigma}\left(0\right)=\left[0,0,0\right]^{\top}$. The design
parameters are $\gamma_{1}=1$, $\gamma_{2}=0.1$, $k_{w}=3$, $\bar{\delta}=\underline{\delta}=1.2$,
$\xi_{0}=1.2$, $\xi_{\infty}=0.04$, and $\ell=4$.

The color notation used in the simulation graphs below is as follows:
green color indicates true value, red refers to the performance of
the semi-direct nonlinear stochastic estimator on $\mathbb{SO}\left(3\right)$
discussed in Subsection \ref{subsec:SO3_PPF_STCH_Passive-Filter},
and blue demonstrates the performance of the direct nonlinear stochastic
estimator on $\mathbb{SO}\left(3\right)$ presented in Subsection
\ref{subsec:SO3_PPF_STCH_Explicit-Filter}. Magenta shows measured
values while orange and black refer to the prescribed performance
response.

\begin{figure}[h!]
	\centering{}\includegraphics[scale=0.26]{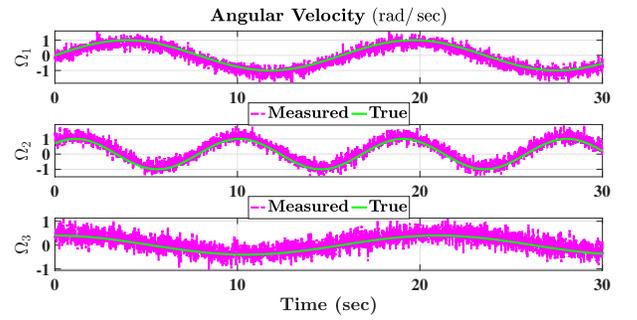}\caption{\label{fig:SO3_PPF_STCH_3}Angular velocities: true and measured values.}
\end{figure}

\begin{figure}[h!]
	\centering{}\includegraphics[scale=0.26]{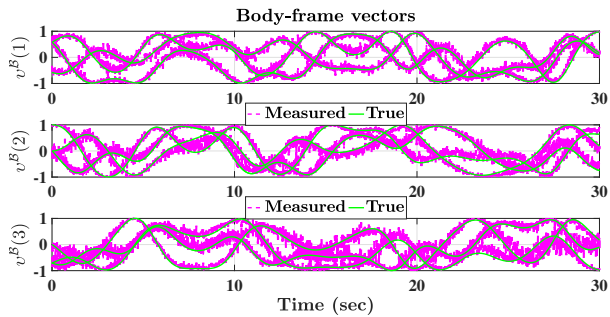}\caption{\label{fig:SO3_PPF_STCH_4}Body-frame vectors: true and measured values.}
\end{figure}

The high levels of noise and bias components associated with the angular
velocity and body-frame measurements are depicted against the true
values in Fig. \ref{fig:SO3_PPF_STCH_3} and Fig. \ref{fig:SO3_PPF_STCH_4}. 

\subsection{Results of Stochastic Estimators in Continuous Form}

\begin{figure}[h!]
	\centering{}\includegraphics[scale=0.26]{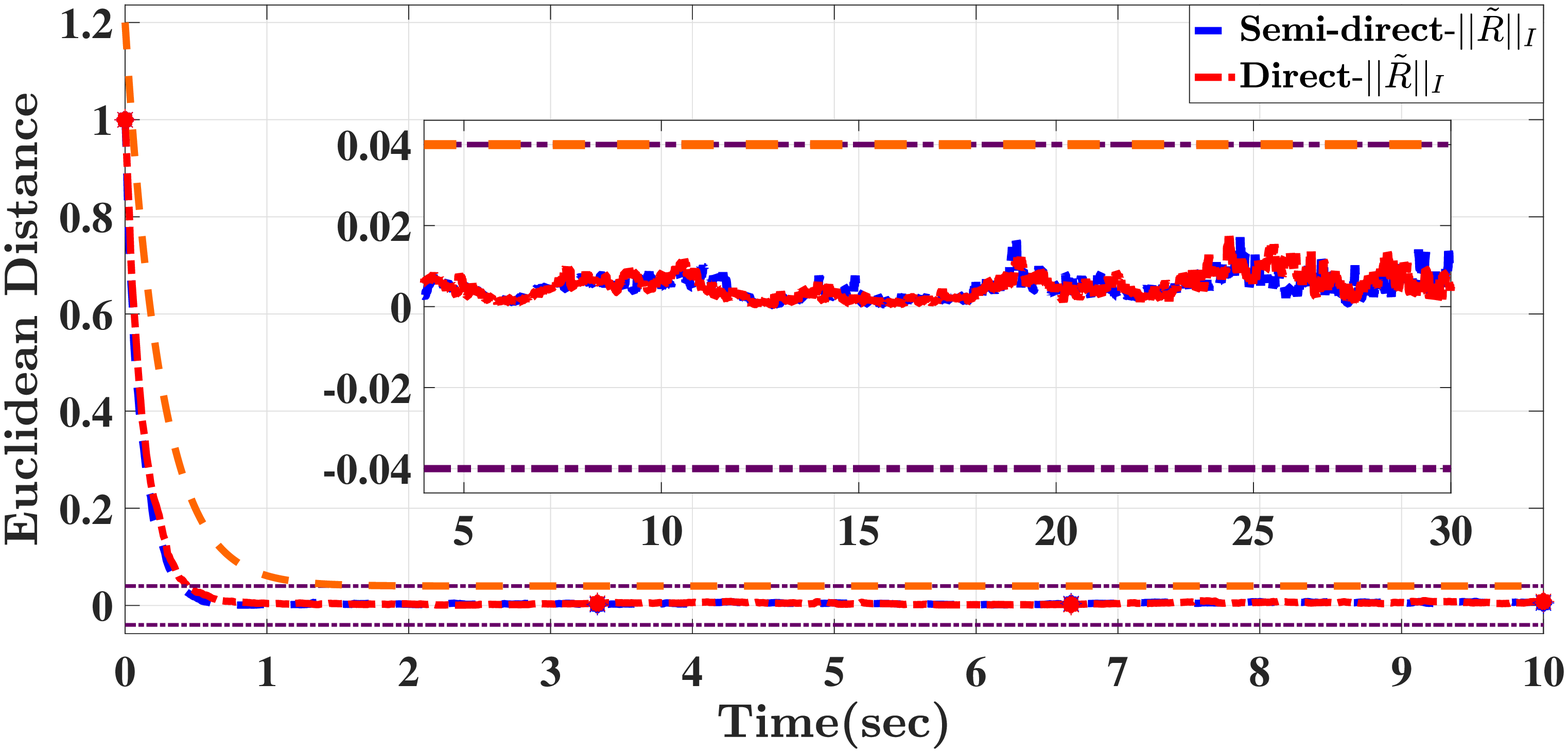}\caption{\label{fig:SO3_PPF_STCH_5}Tracking performance of normalized Euclidean
		distance within PPF.}
\end{figure}

\begin{figure}[h!]
	\centering{}\includegraphics[scale=0.37]{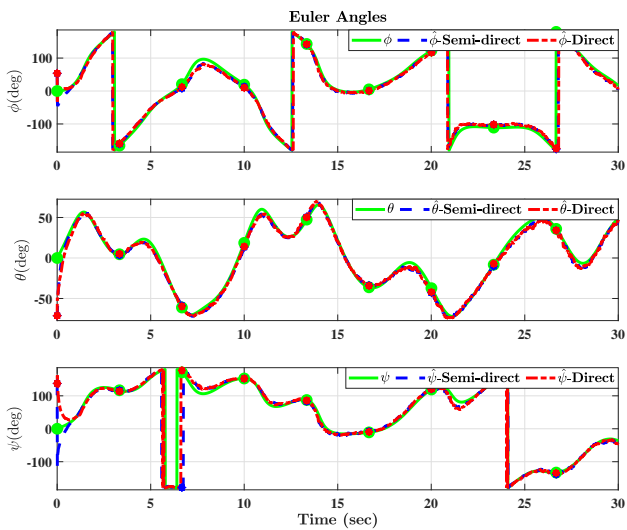}\caption{\label{fig:SO3_PPF_STCH_6}Euler angles: true vs estimate.}
\end{figure}

The systematic and smooth convergence of the normalized Euclidean
distance error $||\tilde{R}||_{I}=\frac{1}{4}{\rm Tr}\{\mathbf{I}_{3}-R^{\top}\hat{R}\}$
from a predetermined large set to a given small residual set is presented
in Fig. \ref{fig:SO3_PPF_STCH_5}. For an explicit illustration of
the systematic convergence, the transient response has been presented
over a period of (0-10) seconds, while the steady-state behavior has
been plotted in a sub-figure of Fig. \ref{fig:SO3_PPF_STCH_5} over
a period of (4-30) seconds. As Fig. \ref{fig:SO3_PPF_STCH_5} shows,
the tracking error started very near to the unstable equilibria within
a given large set and obeyed the dynamic decreasing boundaries of
the PPF. As such, the prescribed performance has been successfully
achieved by utilizing the proposed robust stochastic estimators. The
output performance of the three Euler angle estimates of the proposed
estimators plotted against the true Euler angles are given in Fig.
\ref{fig:SO3_PPF_STCH_6}. It can be noticed that a smooth and fast
tracking performance is achieved in a short period of time. %

Table \ref{tab:SO3_PPF_STCH_1} presents a summary of statistical
details of the error ($||\tilde{R}||_{I}$), namely the mean and STD.
This comparison allows to assess the steady-state error performance
of the proposed estimators examining their oscillatory behavior. It
can be noticed that both estimators have extremely small mean and
STD of $||\tilde{R}||_{I}$. Nonetheless, the semi-direct nonlinear
stochastic attitude estimator with prescribed performance has a remarkably
smaller mean and STD of $||\tilde{R}||_{I}$ in comparison with the
direct nonlinear stochastic estimator with prescribed performance.
Numerical results listed in Table \ref{tab:SO3_PPF_STCH_1} confirm
the robustness of the proposed estimators against large error initialization
and uncertainties in sensor measurements as demonstrated in Fig. \ref{fig:SO3_PPF_STCH_3},
\ref{fig:SO3_PPF_STCH_4}, \ref{fig:SO3_PPF_STCH_5}, and \ref{fig:SO3_PPF_STCH_6}. 

\begin{table}[h!]
	\caption{\label{tab:SO3_PPF_STCH_1}The statistical analysis of the proposed
		estimators.}
	
	\centering{}%
	\begin{tabular}{c|>{\centering}p{2.5cm}|>{\centering}p{2.5cm}}
		\hline 
		\multicolumn{3}{c}{Output data of $||\tilde{R}||_{I}$ over the period (1-30 sec)}\tabularnewline
		\hline 
		\hline 
		Estimator   & Semi-direct   & Direct\tabularnewline
		\hline 
		Mean   & $3.8\times10^{-3}$   & $5.2\times10^{-3}$\tabularnewline
		\hline 
		STD   & $2.1\times10^{-3}$   & $2.6\times10^{-3}$\tabularnewline
		\hline 
	\end{tabular}
\end{table}

\subsection{Results of Stochastic Estimators in Discrete Form}

\begin{figure}[h!]
	\centering{}\includegraphics[scale=0.38]{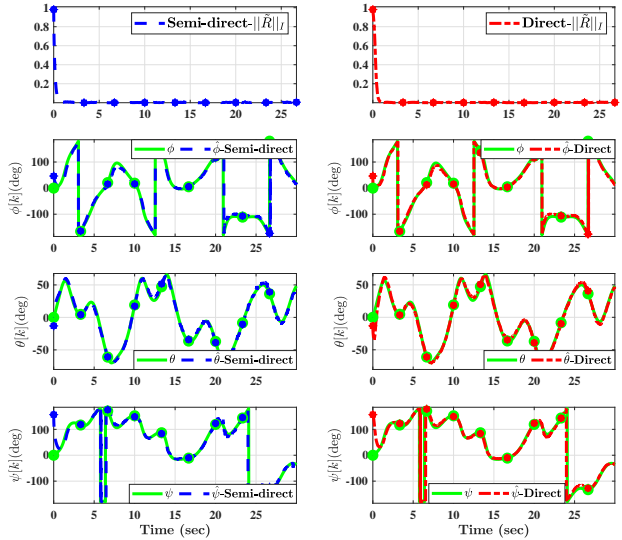}\caption{\label{fig:SO3_PPF_STCH_Disc}Tracking performance of $||\tilde{R}[k]||_{I}$
		and Euler angles of the proposed filters in discrete form.}
\end{figure}

This subsection demonstrates the output performance of the proposed
estimators in their discrete form in Subsection \ref{Subsec:SO3PPF-Discrete}
at small sampling interval. Set the sampling time to $\Delta t=0.01$
seconds. Let the angular velocity and body-frame measurements be analogous
to Fig. \ref{fig:SO3_PPF_STCH_3} and \ref{fig:SO3_PPF_STCH_4}, respectively.
Consider the following definition of the initial attitude (true and
estimate):
\[
R\left(0\right)=\mathbf{I}_{3},\hspace{1em}\hat{R}[0]=\left[\begin{array}{ccc}
-0.8959 & -0.1209 & 0.4275\\
0.3824 & -0.6998 & 0.6034\\
0.2262 & 0.7041 & 0.6731
\end{array}\right]
\]
Let the initial estimates be $\hat{b}\left(0\right)=\left[0,0,0\right]^{\top}$,
$\hat{\sigma}\left(0\right)=\left[0,0,0\right]^{\top}$ , and set
the design parameters to $\gamma_{1}=1$, $\gamma_{2}=0.1$, $k_{w}=3$,
$\bar{\delta}=\underline{\delta}=1.2$, $\xi_{0}=1.2$, $\xi_{\infty}=0.04$,
and $\ell=4$. Fig. \ref{fig:SO3_PPF_STCH_Disc} reveals that the
tracking error of $||\tilde{R}[k]||_{I}$ started very near to the
unstable equilibria and reduced virtually close to the origin. Additionally,
Fig. \ref{fig:SO3_PPF_STCH_Disc} illustrates the output performance
of the Euler angle estimates ($\hat{\phi}[k]$, $\hat{\theta}[k]$,
and $\hat{\psi}[k]$) of the proposed estimators against the true
Euler angles. Overall, Fig. \ref{fig:SO3_PPF_STCH_Disc} demonstrates
smooth and fast tracking performance of the proposed estimators in
a short period of time.

The simulation results reveal the effectiveness and confirm the robustness
of the two proposed nonlinear stochastic estimators against uncertain
measurements and large initialization error. In addition, both estimators
are able to guide the tracking error to comply with the dynamically
decreasing constraints set by the user, and thereby the proposed estimator
design allows to achieve guaranteed measures of transient and steady-state
performance. These advantageous features are not offered by the earlier
proposed attitude estimators such as \cite{mahony2008nonlinear,hamel2006attitude,zlotnik2017nonlinear,grip2012attitude,mahony2005complementary,lee2012exponential}.
It should be noted that the semi-direct nonlinear stochastic attitude
estimator requires attitude reconstruction, and in our case SVD \cite{markley1988attitude}
was employed to obtain $\tilde{R}=R_{y}^{\top}\hat{R}$. Nonetheless,
despite the higher computational power requirement of the semi-direct
estimator, both proposed estimators display remarkable convergence
properties as detailed in Table \ref{tab:SO3_PPF_STCH_1}.

\section{Conclusion\label{sec:SO3_PPF_STCH_Conclusion}}

Successful robotic applications rely heavily on the convergence properties
and steady-state behavior or the attitude estimators. Hence, it is
crucial to account for the fact that attitude estimators are subject
to large error in initialization and uncertainties in measurements,
in particular when low-cost inertial measurement units are utilized
for data collection. This paper addresses the above-mentioned challenge
by proposing two nonlinear stochastic attitude estimators on $\mathbb{SO}\left(3\right)$
with guaranteed performance of transient and steady-state error. Both
estimators reformulate the attitude error in terms of normalized Euclidean
distance and then relax it from its constrained to unconstrained form,
termed transformed error. This allows the transformed error and the
normalized Euclidean distance of the attitude error to be regulated
to the origin in probability from almost any initial condition. In
addition, the equilibrium point has been proven to be independent
of the unknown bias and noise components attached to the angular velocity
measurements. Simulation results demonstrate the ability of the proposed
estimators to obey the predefined characteristics of transient and
steady-state performance set by the user and show their robustness
against high level of uncertainties in the measurements and large
initial attitude error.

\section*{Appendix A\label{sec:SO3_PPF_STCH_AppendixA} }
\begin{center}
	\textbf{\large{}Proof of Lemma \ref{Lemm:SO3_PPF_STCH_1}}{\large\par}
	\par\end{center}

Define the attitude as $R\in\mathbb{SO}\left(3\right)$. Rodriguez
parameters vector $\rho\in\mathbb{R}^{3}$ is employed for attitude
representation such that the related map from vector form to $\mathbb{SO}\left(3\right)$
is given by $\mathcal{R}_{\rho}:\mathbb{R}^{3}\rightarrow\mathbb{SO}\left(3\right)$
\cite{shuster1993survey,hashim2019AtiitudeSurvey} such that
\begin{align}
\mathcal{R}_{\rho}\left(\rho\right)= & \frac{1}{1+||\rho||^{2}}\left((1-||\rho||^{2})\mathbf{I}_{3}+2\rho\rho^{\top}+2\left[\rho\right]_{\times}\right)\label{eq:SO3_PPF_STCH_SO3_Rodr}
\end{align}
with direct substitution of \eqref{eq:SO3_PPF_STCH_SO3_Rodr} in \eqref{eq:SO3_PPF_STCH_Ecul_Dist}
it is straight forward to find
\begin{equation}
||R||_{I}=\frac{||\rho||^{2}}{1+||\rho||^{2}}\label{eq:SO3_PPF_STCH_TR2}
\end{equation}
Similarly, for $\mathcal{R}_{\rho}=\mathcal{R}_{\rho}\left(\rho\right)$
one has
\begin{align*}
\boldsymbol{\mathcal{P}}_{a}\left(R\right)=\frac{1}{2}\left(\mathcal{R}_{\rho}-\mathcal{R}_{\rho}^{\top}\right)= & 2\frac{1}{1+||\rho||^{2}}\left[\rho\right]_{\times}
\end{align*}
such that the vex of the above anti-symmetric operator is governed
by
\begin{equation}
\mathbf{vex}\left(\boldsymbol{\mathcal{P}}_{a}\left(R\right)\right)=2\frac{\rho}{1+||\rho||^{2}}\label{eq:SO3_PPF_STCH_VEX_Pa}
\end{equation}
Hence, from \eqref{eq:SO3_PPF_STCH_TR2} one obtains
\begin{equation}
\left(1-||R||_{I}\right)||R||_{I}=\frac{||\rho||^{2}}{\left(1+||\rho||^{2}\right)^{2}}\label{eq:SO3_PPF_STCH_append1}
\end{equation}
and from \eqref{eq:SO3_PPF_STCH_VEX_Pa} it can be shown that
\begin{equation}
||\mathbf{vex}\left(\boldsymbol{\mathcal{P}}_{a}\left(R\right)\right)||^{2}=4\frac{||\rho||^{2}}{\left(1+||\rho||^{2}\right)^{2}}\label{eq:SO3_PPF_STCH_append2}
\end{equation}
Therefore, \eqref{eq:SO3_PPF_STCH_append1} and \eqref{eq:SO3_PPF_STCH_append2}
justify \eqref{eq:SO3_PPF_STCH_lemm1_1} in Lemma \ref{Lemm:SO3_PPF_STCH_1}.
In Subsection \ref{subsec:SO3_PPF_STCH_Explicit-Filter} it is assumed
that $\sum_{i=1}^{n}s_{i}=3$ which means that ${\rm Tr}\left\{ M^{\mathcal{B}}\right\} =3$.
Since $||M^{\mathcal{B}}R||_{I}=\frac{1}{4}{\rm Tr}\left\{ M^{\mathcal{B}}\left(\mathbf{I}_{3}-R\right)\right\} $,
from angle-axis parameterization in \eqref{eq:SO3_PPF_STCH_att_ang},
one has

\begin{align}
||M^{\mathcal{B}}R||_{I} & =\frac{1}{4}{\rm Tr}\left\{ -M^{\mathcal{B}}\left(\sin(\theta)\left[u\right]_{\times}+\left(1-\cos(\theta)\right)\left[u\right]_{\times}^{2}\right)\right\} \nonumber \\
& =-\frac{1}{4}{\rm Tr}\left\{ M^{\mathcal{B}}\left(1-\cos(\theta)\right)\left[u\right]_{\times}^{2}\right\} \label{eq:SO3_PPF_STCH_append3}
\end{align}
with ${\rm Tr}\left\{ M^{\mathcal{B}}\left[u\right]_{\times}\right\} =0$
as defined in identity \eqref{eq:SO3_PPF_STCH_Identity6}. The following
relation holds \cite{murray1994mathematical} 
\begin{equation}
||R||_{I}=\frac{1}{4}{\rm Tr}\left\{ \mathbf{I}_{3}-R\right\} =\frac{1}{2}\left(1-{\rm cos}\left(\theta\right)\right)={\rm sin}^{2}\left(\frac{\theta}{2}\right)\label{eq:SO3_PPF_STCH_append4}
\end{equation}
such that the relation between Rodriguez parameters vector and angle-axis
parameterization is given by \cite{shuster1993survey} 
\[
u={\rm cot}\left(\theta/2\right)\rho
\]
From identity \eqref{eq:SO3_PPF_STCH_Identity3}, $\left[u\right]_{\times}^{2}=-||u||^{2}\mathbf{I}_{3}+uu^{\top}$.
Therefore, the result in \eqref{eq:SO3_PPF_STCH_append3} becomes
\begin{align*}
||M^{\mathcal{B}}R||_{I} & =\frac{1}{2}||R||_{I}u^{\top}\bar{\mathbf{M}}^{\mathcal{B}}u\\
& =\frac{1}{2}||R||_{I}{\rm cot}^{2}\left(\frac{\theta}{2}\right)\rho^{\top}\bar{\mathbf{M}}^{\mathcal{B}}\rho
\end{align*}
Also, from \eqref{eq:SO3_PPF_STCH_append4}, one has ${\rm cos}^{2}\left(\theta/2\right)=1-||R||_{I}$
which shows that 
\[
{\rm tan}^{2}\left(\theta/2\right)=\frac{||R||_{I}}{1-||R||_{I}}
\]
Accordingly, $||M^{\mathcal{B}}R||_{I}$ can be represented in terms
of Rodriguez parameters vector by 
\begin{align}
||M^{\mathcal{B}}R||_{I} & =\frac{1}{2}(1-||R||_{I})\rho^{\top}\bar{\mathbf{M}}^{\mathcal{B}}\rho\nonumber \\
& =\frac{1}{2}\frac{\rho^{\top}\bar{\mathbf{M}}^{\mathcal{B}}\rho}{1+||\rho||^{2}}\label{eq:SO3_PPF_STCH_append_MBR_I}
\end{align}
With aid of identity \eqref{eq:SO3_PPF_STCH_Identity1} and \eqref{eq:SO3_PPF_STCH_Identity4},
the anti-symmetric projection operator is expressed in the sense of
Rodriquez parameters vector: 
\begin{align*}
\boldsymbol{\mathcal{P}}_{a}(M^{\mathcal{B}}R)= & \frac{M^{\mathcal{B}}\rho\rho^{\top}-\rho\rho^{\top}M^{\mathcal{B}}+M^{\mathcal{B}}\left[\rho\right]_{\times}+\left[\rho\right]_{\times}M^{\mathcal{B}}}{1+||\rho||^{2}}\\
= & \frac{\left[\left({\rm Tr}\left\{ M^{\mathcal{B}}\right\} \mathbf{I}_{3}-M^{\mathcal{B}}+\left[\rho\right]_{\times}M^{\mathcal{B}}\right)\rho\right]_{\times}}{1+||\rho||^{2}}
\end{align*}
such that the vex operator of the result above is equivalent to
\begin{align}
\mathcal{\mathbf{vex}}\left(\boldsymbol{\mathcal{P}}_{a}(M^{\mathcal{B}}R)\right) & =\frac{1}{1+||\rho||^{2}}\left(\mathbf{I}_{3}-\left[\rho\right]_{\times}\right)\bar{\mathbf{M}}^{\mathcal{B}}\rho\label{eq:SO3_PPF_STCH_append_MBR_VEX}
\end{align}
with the 2-norm of \eqref{eq:SO3_PPF_STCH_append_MBR_VEX} being equal
to 
\begin{align*}
||\mathbf{vex}\left(\boldsymbol{\mathcal{P}}_{a}(M^{\mathcal{B}}R)\right)||^{2} & =\frac{\rho^{\top}\bar{\mathbf{M}}^{\mathcal{B}}\left(\mathbf{I}_{3}-\left[\rho\right]_{\times}^{2}\right)\bar{\mathbf{M}}^{\mathcal{B}}\rho}{\left(1+||\rho||^{2}\right)^{2}}
\end{align*}
From identity \eqref{eq:SO3_PPF_STCH_Identity3} $\left[\rho\right]_{\times}^{2}=-||\rho||^{2}\mathbf{I}_{3}+\rho\rho^{\top}$,
which means that
\begin{align}
||\mathbf{vex}\left(\boldsymbol{\mathcal{P}}_{a}(M^{\mathcal{B}}R)\right)||^{2} & =\frac{\rho^{\top}\bar{\mathbf{M}}^{\mathcal{B}}\left(\mathbf{I}_{3}-\left[\rho\right]_{\times}^{2}\right)\bar{\mathbf{M}}^{\mathcal{B}}\rho}{\left(1+||\rho||^{2}\right)^{2}}\nonumber \\
& =\frac{\rho^{\top}\left(\bar{\mathbf{M}}^{\mathcal{B}}\right)^{2}\rho}{1+||\rho||^{2}}-\frac{\left(\rho^{\top}\bar{\mathbf{M}}^{\mathcal{B}}\rho\right)^{2}}{\left(1+||\rho||^{2}\right)^{2}}\nonumber \\
& \geq\underline{\lambda}\left(1-\frac{\left\Vert \rho\right\Vert ^{2}}{1+\left\Vert \rho\right\Vert ^{2}}\right)\frac{\rho^{\top}\bar{\mathbf{M}}^{\mathcal{B}}\rho}{1+||\rho||^{2}}\label{eq:SO3_PPF_STCH_append_VEX_ineq}
\end{align}
with $\underline{\lambda}=\underline{\lambda}(\bar{\mathbf{M}}^{\mathcal{B}})$
being the minimum singular value of $\bar{\mathbf{M}}^{\mathcal{B}}$
and $\left\Vert R\right\Vert _{I}=\frac{\left\Vert \rho\right\Vert ^{2}}{1+\left\Vert \rho\right\Vert ^{2}}$
as stated in \eqref{eq:SO3_PPF_STCH_TR2}. One has
\begin{align}
1-\left\Vert R\right\Vert _{I} & ={\rm Tr}\left\{ \frac{1}{12}\mathbf{I}_{3}+\frac{1}{4}R\right\} \nonumber \\
& ={\rm Tr}\left\{ \frac{1}{12}\mathbf{I}_{3}+\frac{1}{4}\left(M^{\mathcal{B}}\right)^{-1}M^{\mathcal{B}}R\right\} \label{eq:SO3_PPF_STCH_append_rho2}
\end{align}
Thus, from \eqref{eq:SO3_PPF_STCH_append_VEX_ineq} and \eqref{eq:SO3_PPF_STCH_append_rho2}
the following inequality holds 
{\small
\begin{align*}
||\mathbf{vex}\left(\boldsymbol{\mathcal{P}}_{a}\left(M^{\mathcal{B}}R\right)\right)||^{2} & \geq\frac{\underline{\lambda}}{2}\left(1+{\rm Tr}\left\{ \left(M^{\mathcal{B}}\right)^{-1}M^{\mathcal{B}}R\right\} \right)\left\Vert M^{\mathcal{B}}R\right\Vert _{I}
\end{align*}
}
this proves \eqref{eq:SO3_PPF_STCH_lemm1_3} in Lemma \ref{Lemm:SO3_PPF_STCH_1}.

\section*{Appendix B\label{sec:SO3_PPF_STCH_AppendixB} }
\begin{center}
	\textbf{\large{}{}{}{}{}{}{}{}{}{}{}{}Quaternion Representation}{\large{}{}{}
	} 
	\par\end{center}

\noindent Define $Q=[q_{0},q^{\top}]^{\top}\in\mathbb{S}^{3}$ as
a unit-quaternion with $q_{0}\in\mathbb{R}$ and $q\in\mathbb{R}^{3}$
such that $\mathbb{S}^{3}=\{\left.Q\in\mathbb{R}^{4}\right|||Q||=\sqrt{q_{0}^{2}+q^{\top}q}=1\}$.
$Q^{-1}=[\begin{array}{cc}
q_{0} & -q^{\top}\end{array}]^{\top}\in\mathbb{S}^{3}$ denotes the inverse of $Q$. Define $\odot$ as a quaternion product
where the quaternion multiplication of $Q_{1}=[\begin{array}{cc}
q_{01} & q_{1}^{\top}\end{array}]^{\top}\in\mathbb{S}^{3}$ and $Q_{2}=[\begin{array}{cc}
q_{02} & q_{2}^{\top}\end{array}]^{\top}\in\mathbb{S}^{3}$ is $Q_{1}\odot Q_{2}=[q_{01}q_{02}-q_{1}^{\top}q_{2},q_{01}q_{2}+q_{02}q_{1}+[q_{1}]_{\times}q_{2}]^{\top}$.
The mapping from unit-quaternion ($\mathbb{S}^{3}$) to $\mathbb{SO}\left(3\right)$
is described by $\mathcal{R}_{Q}:\mathbb{S}^{3}\rightarrow\mathbb{SO}\left(3\right)$
\begin{align}
\mathcal{R}_{Q} & =(q_{0}^{2}-||q||^{2})\mathbf{I}_{3}+2qq^{\top}+2q_{0}\left[q\right]_{\times}\in\mathbb{SO}\left(3\right)\label{eq:NAV_Append_SO3}
\end{align}
The quaternion identity is described by $Q_{{\rm I}}=[\pm1,0,0,0]^{\top}$
with $\mathcal{R}_{Q_{{\rm I}}}=\mathbf{I}_{3}$. For more details visit \cite{hashim2019AtiitudeSurvey}. Define the estimate
of $Q=[q_{0},q^{\top}]^{\top}\in\mathbb{S}^{3}$ as $\hat{Q}=[\hat{q}_{0},\hat{q}^{\top}]^{\top}\in\mathbb{S}^{3}$
with $\mathcal{R}_{\hat{Q}}=(\hat{q}_{0}^{2}-||\hat{q}||^{2})\mathbf{I}_{3}+2\hat{q}\hat{q}^{\top}+2\hat{q}_{0}\left[\hat{q}\right]_{\times}$,
see the map in \eqref{eq:NAV_Append_SO3}. The equivalent quaternion
representation of the filter in \eqref{eq:SO3_PPF_STCH_Rest_dot_Ry},
\eqref{eq:SO3_PPF_STCH_b_est_Ry}, \eqref{eq:SO3_PPF_STCH_s_est_Ry},
and \eqref{eq:SO3_PPF_STCH_W_Ry} is:
\[
\begin{cases}
\left[\begin{array}{c}
0\\
\upsilon_{i}^{\mathcal{B}}
\end{array}\right] & =Q^{-1}\odot\left[\begin{array}{c}
0\\
\upsilon_{i}^{\mathcal{I}}
\end{array}\right]\odot Q\\
Q_{y}: & \text{Reconstructed by QUEST algorithm}\\
\tilde{Q} & =[\tilde{q}_{0},\tilde{q}^{\top}]^{\top}=Q_{y}^{-1}\odot\hat{Q}\\
||\tilde{R}||_{I} & =1-\tilde{q}_{0}^{2}\\
\mu & =\frac{\exp\left(2\mathcal{E}\right)+\exp\left(-2\mathcal{E}\right)+2}{8\xi\bar{\delta}}\\
\Gamma & =\Omega_{m}-\hat{b}-W\\
\dot{\hat{Q}} & =\frac{1}{2}\left[\begin{array}{cc}
0 & -\Gamma^{\top}\\
\Gamma & -\left[\Gamma\right]_{\times}
\end{array}\right]\hat{Q}\\
\dot{\hat{b}} & =2\gamma_{1}\left(\mathcal{E}+1\right)\exp\left(\mathcal{E}\right)\mu\tilde{q}_{0}\tilde{q}\\
\mathcal{\dot{\hat{\sigma}}} & =4\gamma_{2}\left(\mathcal{E}+2\right)\exp\left(\mathcal{E}\right)\mu^{2}\tilde{q}_{0}^{2}{\rm diag}\left(\tilde{q}\right)\tilde{q}\\
W & =2\tilde{q}_{0}\frac{\mathcal{E}+2}{\mathcal{E}+1}\mu{\rm diag}\left(\tilde{q}\right)\hat{\sigma}+2\frac{k_{w}\mathcal{E}\mu-\dot{\xi}/4\xi}{\tilde{q}_{0}}\tilde{q}
\end{cases}
\]

\noindent The equivalent quaternion representation of the filter in
\eqref{eq:SO3_PPF_STCH_Rest_dot_VM}, \eqref{eq:SO3_PPF_STCH_b_est_VM},
\eqref{eq:SO3_PPF_STCH_s_est_VM}, and \eqref{eq:SO3_PPF_STCH_W_VM}
is:{\small
\[
\begin{cases}
\left[\begin{array}{c}
0\\
\upsilon_{i}^{\mathcal{B}}
\end{array}\right] & =Q^{-1}\odot\left[\begin{array}{c}
0\\
\upsilon_{i}^{\mathcal{I}}
\end{array}\right]\odot Q\\
\left[\begin{array}{c}
0\\
\hat{\upsilon}_{i}^{\mathcal{B}}
\end{array}\right] & =\hat{Q}^{-1}\odot\left[\begin{array}{c}
0\\
\upsilon_{i}^{\mathcal{I}}
\end{array}\right]\odot\hat{Q}\\
\boldsymbol{\Upsilon} & =\sum_{i=1}^{n}\frac{s_{i}}{2}\hat{\upsilon}_{i}^{\mathcal{B}}\times\upsilon_{i}^{\mathcal{B}}\\
||M^{\mathcal{B}}\tilde{R}||_{I} & =0.25{\rm Tr}\{M^{\mathcal{B}}-\sum_{i=1}^{n}s_{i}\upsilon_{i}^{\mathcal{B}}\left(\hat{\upsilon}_{i}^{\mathcal{B}}\right)^{\top}\}\\
\boldsymbol{\mathcal{J}} & ={\rm Tr}\left\{ \left(\sum_{i=1}^{n}s_{i}\upsilon_{i}^{\mathcal{B}}\left(\upsilon_{i}^{\mathcal{B}}\right)^{\top}\right)^{-1}\sum_{i=1}^{n}s_{i}\upsilon_{i}^{\mathcal{B}}\left(\hat{\upsilon}_{i}^{\mathcal{B}}\right)^{\top}\right\} \\
\mu & =\frac{\exp\left(2\mathcal{E}\right)+\exp\left(-2\mathcal{E}\right)+2}{8\xi\bar{\delta}}\\
\Gamma & =\Omega_{m}-\hat{b}-W\\
\dot{\hat{Q}} & =\frac{1}{2}\left[\begin{array}{cc}
0 & -\Gamma^{\top}\\
\Gamma & -\left[\Gamma\right]_{\times}
\end{array}\right]\hat{Q}\\
\dot{\hat{b}} & =\gamma_{1}\mu\left(\mathcal{E}+1\right)\exp\left(\mathcal{E}\right)\boldsymbol{\Upsilon}\\
\mathcal{\dot{\hat{\sigma}}} & =\gamma_{2}\left(\mathcal{E}+2\right)\exp\left(\mathcal{E}\right)\mu^{2}{\rm diag}\left(\boldsymbol{\Upsilon}\right)\boldsymbol{\Upsilon}\\
W & =2\frac{\mathcal{E}+2}{\mathcal{E}+1}\mu{\rm diag}\left(\boldsymbol{\Upsilon}\right)\hat{\sigma}+\frac{4}{\underline{\lambda}}\frac{k_{w}\mu\mathcal{E}-\dot{\xi}/\xi}{1+\boldsymbol{\mathcal{J}}}\boldsymbol{\Upsilon}
\end{cases}
\]
}

\section*{Acknowledgment}

The authors would like to thank \textbf{Maria Shaposhnikova} for proofreading
the article.

\bibliographystyle{IEEEtran}
\bibliography{bib_PPF_SO3}
\vspace{220pt}


\end{document}